\pdfoutput=0
\documentclass[11pt]{article}
\usepackage{axodraw}
\usepackage{epsfig}
\usepackage{amsfonts}
\usepackage{amsmath}
\usepackage{bbm,bm}
\usepackage{cite}
\allowdisplaybreaks[1]

\input pix.sty

\newcommand{\la}[1]{\label{#1}}
\newcommand{\be}{\begin{equation}}
\newcommand{\ee}{\end{equation}}
\newcommand{\ba}{\begin{eqnarray}}
\newcommand{\ea}{\end{eqnarray}}
\newcommand{\rmi}[1]{{\mbox{\scriptsize #1}}}
\newcommand{\fig}{fig.~}
\newcommand{\figs}{figs.~}
\newcommand{\eq}{eq.~}
\newcommand{\eqs}{eqs.~}
\newcommand{\se}{sec.~}

\newcommand{\nr}[1]{(\ref{#1})}
\newcommand{\tr}{{\rm Tr\,}}
\newcommand{\nn}{\nonumber \\}
\newcommand{\fr}[2]{{\frac{#1}{#2}}}
\newcommand{\msbar}{{\overline{\mbox{\rm MS}}}}

\newcommand{\Ntau}{N_\tau^{ }}

\renewcommand{\vec}[1]{{\bf #1}}

\newcommand{\tinymsbar}{{\overline{\mbox{\tiny\rm{MS}}}}}
\newcommand{\Lambdamsbar}{{\Lambda_\tinymsbar}}
\newcommand{\alphas}{\alpha_{\rm s}}

\newcommand{\T}{\rmii{$T$}}
\newcommand{\Nf}{N_{\rm f}}
\newcommand{\Nc}{N_{\rm c}}
\newcommand{\Ns}{N_{\rm s}}
\newcommand{\Tc}{T_{\rm c}}
\newcommand{\rO}{r^{ }_0}

\newcommand{\Nt}{N^{ }_\tau}

\newcommand{\E}{\rmii{$E$}}
\renewcommand{\B}{\rmii{$B$}}
\newcommand{\iB}{{\mbox{\tiny{$\scriptstyle{B}$}}}}
\newcommand{\I}{\mathcal{I}}
\newcommand{\J}{\mathcal{J}}
\newcommand{\CF}{C_\rmii{F}}
\newcommand{\gB}{g_\rmii{B}}

\newcommand{\gammaE}{\gamma_\rmii{E}}

\newcommand{\rmO}{{\mathcal{O}}}
\newcommand{\bmu}{\bar\mu}
\newcommand{\CA}{\Nc} 

\def\lsi{\raise0.3ex\hbox{$<$\kern-0.75em\raise-1.1ex\hbox{$\sim$}}}
\def\gsi{\raise0.3ex\hbox{$>$\kern-0.75em\raise-1.1ex\hbox{$\sim$}}}
\newcommand{\lsim}{\mathop{\lsi}}
\newcommand{\gsim}{\mathop{\gsi}}

\newcommand{\nF}{n_\rmii{F}}
\newcommand{\nB}{n_\rmii{B}}
\newcommand{\tlo}{\bar\tau^{ }_\rmi{low}}
\newcommand{\rmii}[1]{{\mbox{\tiny\rm{#1}}}}
\newcommand{\rmiii}[1]{{\mbox{\tiny{$\scriptstyle{\rm#1}$}}}}
\newcommand{\re}{\mathop{\mbox{Re}}}
\newcommand{\im}{\mathop{\mbox{Im}}}
\newcommand{\Tint}[1]{{\hbox{$\sum$}\!\!\!\!\!\!\!\int\,}_{\!\!\!\!\raise-0.9ex\hbox{$\scriptstyle{#1}$}}}
\newcommand{\Tinti}[1]{{{\Sigma}\!\!\!\!\raise0.3ex\hbox{$\int$}_\rmii{${#1}$}}}
\newcommand{\Tintip}[1]{{{\Sigma'}\!\!\!\!\!\raise0.3ex\hbox{$\int$}_\rmii{${#1}$}}}

\newcommand{\bi}{\begin{itemize}}
\newcommand{\ei}{\end{itemize}}
\newcommand{\hide}[1]{ }

\newcommand{\deltabar}{\raise-0.02em\hbox{$\bar{}$}\hspace*{-0.8mm}{\delta}}

\newcommand{\picu}[1]{\;\parbox[c]{30pt}{\begin{picture}(30,30)(0,0)
\SetWidth{1.0}\SetScale{1.0} #1 \end{picture}}\; }
\def\EleA{\picu{%
 \CArc(15,15)(15,0,360)%
 \Lgl(15,0)(15,30)%
 \COval(15,0)(2,2)(0){Black}{Black}%
 \COval(15,30)(2,2)(0){Black}{Black}%
}}
\def\EleB{\picu{%
 \CArc(15,15)(15,0,360)%
 \Lgl(15,0)(15,30)%
 \COval(15,0)(2,2)(0){Black}{Black}%
 \COval(15,30)(2,2)(0){Black}{Black}%
 \Agl(29,15)(8,100,260)%
}}
\def\EleBB{\picu{%
 \CArc(15,15)(15,0,360)%
 \Lgl(15,0)(15,30)%
 \Lgl(0,15)(12,15)%
 \Lgl(18,15)(30,15)%
 \COval(15,0)(2,2)(0){Black}{Black}%
 \COval(15,30)(2,2)(0){Black}{Black}%
}}
\def\EleC{\picu{%
 \CArc(15,15)(15,0,360)%
 \Agl(7,4)(8,-30,130)%
 \Agl(23,26)(8,150,310)%
 \COval(15,0)(2,2)(0){Black}{Black}%
 \COval(15,30)(2,2)(0){Black}{Black}%
}}
\def\EleD{\picu{%
 \CArc(15,15)(15,0,360)%
 \Agl(23,4)(8,50,210)%
 \Agl(23,26)(8,150,310)%
 \COval(15,0)(2,2)(0){Black}{Black}%
 \COval(15,30)(2,2)(0){Black}{Black}%
}}
\def\EleE{\picu{%
 \CArc(15,15)(15,0,360)%
 \Agl(40,30)(25,180,242)%
 \Agl(40,0)(25,118,137)%
 \Agl(40,0)(25,150,180)%
 \COval(15,0)(2,2)(0){Black}{Black}%
 \COval(15,30)(2,2)(0){Black}{Black}%
}}
\def\EleF{\picu{%
 \CArc(15,15)(15,0,360)%
 \Lgl(15,0)(15,30)%
 \Agl(40,40)(26,205,245)%
 \COval(15,0)(2,2)(0){Black}{Black}%
 \COval(15,30)(2,2)(0){Black}{Black}%
}}
\def\EleG{\picu{%
 \CArc(15,15)(15,0,360)%
 \Agl(40,15)(30,150,210)%
 \Agl(-10,15)(30,-30,30)%
 \COval(15,0)(2,2)(0){Black}{Black}%
 \COval(15,30)(2,2)(0){Black}{Black}%
}}
\def\EleH{\picu{%
 \CArc(15,15)(15,0,360)%
 \Lgl(15,0)(15,12)%
 \Agl(10,21.5)(10,-65,65)%
 \Agl(20,21.5)(10,115,245)%
 \COval(15,0)(2,2)(0){Black}{Black}%
 \COval(15,30)(2,2)(0){Black}{Black}%
}}
\def\EleI{\picu{%
 \CArc(15,15)(15,0,360)%
 \Lgl(15,0)(15,30)%
 \Lgl(15,15)(30,15)%
 \COval(15,0)(2,2)(0){Black}{Black}%
 \COval(15,30)(2,2)(0){Black}{Black}%
}}
\def\EleJ{\picu{%
 \CArc(15,15)(15,0,360)%
 \Lgl(15,0)(15,30)%
 \COval(15,0)(2,2)(0){Black}{Black}%
 \COval(15,30)(2,2)(0){Black}{Black}%
 \GCirc(15,15){4}{0.5}
}}

\makeatletter \@addtoreset{equation}{section} \makeatother
\renewcommand{\theequation}{\arabic{section}.\arabic{equation}}
\makeatletter
\renewcommand\section{\@startsection {section}{1}{\z@}%
                                   {-5.5ex \@plus -1ex \@minus -.2ex}
                                   {2.3ex \@plus.2ex}%
                                   {\normalfont\large\bfseries}}
\renewcommand\subsection{\@startsection{subsection}{2}{\z@}%
                                     {-3.25ex\@plus -1ex \@minus -.2ex}%
                                     {1.5ex \@plus .2ex}%
                                     {\normalfont\normalsize\bfseries}}
\renewcommand\thesection {\@arabic\c@section}
\renewcommand\thesubsection   {\thesection.\@arabic\c@subsection}
\renewcommand{\@seccntformat}[1]{%
\csname the#1\endcsname.\hspace{1.0em}}
\makeatother

\begin{document}

\flushbottom


\begin{titlepage}

\begin{flushright}
August 2022
\end{flushright}

\begin{centering}
\vfill

{\Large{\bf
 Lattice study of a magnetic contribution \\[2mm]
 ~to heavy quark momentum diffusion
}} 

\vspace{0.8cm}

D.~Banerjee$^{\rmi{a},\rmi{b}}_{ }$, 
S.~Datta$^{\rmi{c}}_{ }$, 
M.~Laine$^{\rmi{d}}_{ }$
 
\vspace{0.8cm}

$^\rmi{a}$%
{\em
Saha Institute of Nuclear Physics, \\
1/AF Bidhannagar, Kolkata 700064, India \\}

\vspace*{0.3cm}

$^\rmi{b}$%
{\em
Homi Bhabha National Institute, Training School Complex, \\ 
Anushaktinagar, Mumbai 400094, India \\}

\vspace*{0.3cm}

$^\rmi{c}$%
{\em
Department of Theoretical Physics, 
Tata Institute of Fundamental Research, \\  
Homi Bhabha Road, Mumbai 400005, India \\}

\vspace*{0.3cm}

$^\rmi{d}$%
{\em
AEC, 
Institute for Theoretical Physics, 
University of Bern, \\ 
Sidlerstrasse 5, CH-3012 Bern, Switzerland \\}

\vspace*{0.8cm}

\mbox{\bf Abstract}
 
\end{centering}

\vspace*{0.3cm}
 
\noindent
Heavy quarks placed within a hot QCD medium undergo Brownian motion, 
characterized by specific transport coefficients. Their determination 
can be simplified by expanding them in $T/M$, 
where $T$ is the temperature and $M$ is a heavy quark mass.
The leading term in the expansion
originates from the colour-electric part of a Lorentz force, 
whereas the next-to-leading order involves the colour-magnetic part. 
We measure a colour-magnetic 2-point correlator in quenched QCD
at $T \sim  (1.2 - 2.0) \Tc^{ }$. Employing 
multilevel techniques and non-perturbative renormalization, a good 
signal is obtained, and its continuum extrapolation can be estimated. 
Modelling the shape of the corresponding spectral function,
we subsequently extract the momentum diffusion coefficient, $\kappa$. 
For charm (bottom) quarks, the magnetic contribution 
adds $\sim 30\%$ ($10\%$) to the electric one. The same
increases apply also to the drag coefficient, $\eta$.
As an aside, the colour-magnetic spectral function is computed at NLO.

\vfill

 
\end{titlepage}

\tableofcontents

%
\section{Introduction}

Thanks to the inertia provided by their mass, 
heavy quarks tend to interact
less efficiently 
with a hot QCD medium than light quarks or gluons. 
This turns them into a tractable probe for the properties of the  
strongly interacting plasma that is 
generated in heavy ion collision experiments, 
a fact that has inspired theoretical 
and phenomenological investigations of a broad variety   
(cf.,\ e.g.,\ ref.~\cite{pheno} for a review). 

A quantitative study of the movement of heavy quarks within 
an expanding hydrodynamical background can be based on the 
Langevin equation~\cite{bs,mt}, 
\be
 \dot{p}^{ }_i(t) =  - \eta \, p^{ }_i(t) + \xi^{ }_i(t)
 \;, \quad 
 \langle\, \xi^{ }_i(t) \,\rangle = 0 
 \;, \quad 
 \langle\, \xi^{ }_i(t') \, \xi^{ }_j(t) \,\rangle = 
 \kappa \, \delta^{ }_{ij} \, \delta(t - t')
 \;, \la{langevin} 
\ee
where the $p^{ }_i$ denote components of momenta. 
The magnitude of the kicks that the medium
exerts are incorporated in the autocorrelator
of the stochastic noise, parametrized by 
the momentum diffusion coefficient, $\kappa$.
Thanks to a fluctuation-dissipation relation
(cf.\ \eq\nr{eta}), 
$\kappa$ in turn
determines the drag coefficient, $\eta$, which 
according to \eq\nr{langevin} can be  
interpreted as a kinetic equilibration rate. 

A QCD-based 
determination of $\kappa$ can be simplified
by viewing it as an expansion in $T/M$, where $T$ is the 
temperature and $M$ is a heavy quark mass. The leading term, 
which remains present in the asymptotic limit of a very small $T/M$, 
originates from a colour-electric two-point correlator~\cite{cst,kappaE}.  
Let us denote this term by $\kappa^{ }_{\E}$. 
Many lattice studies of $\kappa^{ }_{\E}$ 
have been reported in quenched 
QCD~\cite{lat2,lat25,lat3,lat4,lat5,lat6,lat6b}, 
and even though further improvements 
are desirable, 
related for instance to non-perturbative renormalization, 
unquenching, and analytic continuation, 
the patterns that have emerged so far 
point towards a consistent picture. 

In view of the fact that the charm quark is not extremely heavy  
it can be asked, 
however, 
how large the corrections from a finite $T/M$ could be. 
It turns out that there are 
a number of separate contributions at this order, and a particularly
important one originates from the magnetic part of a coloured
Lorentz force~\cite{1overM}. 
The colour-magnetic contribution experiences  non-trivial 
renormalization~\cite{Bmatch}, but otherwise its lattice 
determination should not be harder 
than for the colour-electric case. 
Indeed attempts in this direction have been
initiated by several groups recently~\cite{new1,new2}.

The goal of the present paper is to study the colour-magnetic 
contribution to heavy quark momentum diffusion through moderate-scale
lattice simulations. On the theoretical side, a significant part of the 
effort goes to clarifying the renormalization of the observable, which
we aim to accomplish (partly) on the non-perturbative level. On the 
physics side, different orders in $T/M$ permit for
interesting comparisons of the charm and bottom quark properties. 

Our presentation is organized as follows. 
After formulating the basic framework
(cf.\ \se\ref{se:framework}), 
we specify its lattice implementation and 
the set of simulations carried out
(cf.\ \se\ref{se:methods}). 
Data analysis is split into two parts, 
an in principle well-defined continuum extrapolation 
in the imaginary-time domain
(cf.\ \se\ref{se:data}), 
as well as a spectral analysis in the
Minkowskian one, which is necessarily 
of a more exploratory nature
(cf.\ \se\ref{se:implications}). 
In spite of the  
systematic uncertainties of the latter step, 
the results show a coherent pattern, and permit for 
us to formulate physical 
conclusions (cf.\ \se\ref{se:concl}). 
Technical details, related to non-perturbative renormalization,   
tree-level improvement, 
next-to-leading order corrections to the colour-magnetic 
spectral function, 
and fitting strategies, 
are relegated to appendices A, B, C and D, respectively. 

%
\section{Basic framework}
\la{se:framework}

The correlation function that we are concerned with can formally
be expressed as~\cite{1overM} 
\be
 \bigl[ G^{ }_{\B}(\tau) \bigr]^{ }_\rmi{bare}
 \; \equiv \;
 \frac{
        \sum_i \re\tr \bigl \langle 
                      U({1}/{T};\tau)\, 
                        \bigl[ gB^{ }_i(\tau) \bigr]^{ }_\rmi{bare} \, 
                      U(\tau;0) \, 
                        \bigl[ gB^{ }_i(0) \bigr]^{ }_\rmi{bare} \, 
                      \bigr \rangle
      }{
        3 \re\tr\langle U({1}/{T};0) \rangle
      }
 \;. \la{GB_def}
\ee
Here 
$\tau \in (0,1/T)$ is an imaginary-time coordinate; 
$U(\tau^{ }_2;\tau^{ }_1)$
is a straight Wilson line in the time direction; 
and $g B^{ }_i$ represents a colour-magnetic field strength, defined
as a Hermitean matrix transforming in the adjoint representation. 

In order to measure \eq\nr{GB_def} on the lattice (L), 
we make use of the Wilson gauge action. 
The magnetic field is defined as 
\be
 \bigl[ g B^{ }_i \bigr]^{ }_{\rmi{bare},\rmii{L}} \; \equiv \;  
 \frac{\sum_{j,k}\epsilon^{ }_{ijk} \widehat{F}^{ }_{jk}}{2 \imath }
 \;, \la{B_bare}
\ee
where $\imath$ is the imaginary unit and
$ 
 \widehat{F}^{ }_{jk}
$ 
is a field strength obtained from a clover,   
\ba
 \widehat{F}^{ }_{jk}(x) & \equiv & 
 \frac{  Q^{ }_{jk}(x) - Q^{ }_{kj}(x)  }{8a^2} 
 \;, \la{clover} \\[2mm] 
 Q^{ }_{jk}(x) & \equiv & 
 P^{ }_{jk}(x) + P^{ }_{k-j}(x) + P^{ }_{-j-k}(x) + P^{ }_{-k j}(x)
 \;, \la{plaquette}
\ea
where 
$
 P^{ }_{jk}(x) \; \equiv \;  
 U^{ }_{j}(x)\, U^{ }_k (x + a \,\hat{\!j})\, 
 U^{\dagger}_{j}(x + a \hat{k})\, U^\dagger_k(x)
$ 
is a plaquette; 
$U^{ }_j$ is a link matrix pointing in the $j$-direction;
$U^{ }_{-j}(x) \equiv U^\dagger_{j}(x - a \,\hat{\!j})$;
$\,\hat{\!j}$ is a unit vector; and $a$ is the lattice spacing.

At loop level,
the colour-magnetic field experiences 
non-trivial renormalization~\cite{hqet1}. 
In dimensional regularization (DR), this means that the bare 
correlator needs to be multiplied by a renormalization factor
in order to obtain a finite $\msbar$ correlator, 
\be
 \bigl[ G^{ }_{\B}(\tau) \bigr]^{ }_\rmi{renorm,$\bmu$}
 = 
 \biggl[
   1 + \frac{g^2 \mu^{-2\epsilon}_{ }}{(4\pi)^2}
       \frac{ \Nc^{ } }{\epsilon}
     + \rmO(g^4) 
 \biggr]^2 
 \, 
 \bigl[ G^{ }_{\B}(\tau) \bigr]^{ }_{\rmi{bare},\rmii{DR}}
 \;, \la{msbar}
\ee
where $g^2 = 4\pi\alphas$ is a renormalized coupling, 
$\mu$ is a scale parameter, 
$\Nc^{ }= 3$, 
and the space-time dimension has been written as $D = 4 - 2\epsilon$.
Because of this renormalization, the correlator now depends on the $\msbar$
renormalization scale, 
$
  \bmu^2 \equiv 4\pi \mu^2 e^{ - \gammaE }_{ }
$. 
In physical results, the correlator is 
multiplied by a Wilson coefficient, which cancels the scale dependence. 

Considering the Lorentz force that acts on heavy quarks, 
and matching a full thermal QCD computation onto a static effective
theory one, the Wilson coefficient relevant for the colour-magnetic
correlator was determined in ref.~\cite{Bmatch}. Concretely, we can write
\ba
 \bigl[ G^{ }_{\B}(\tau) \bigr]^{ }_\rmi{physical}
 & = & 
 c^{2}_{\B}(\bmu) \, 
 \bigl[ G^{ }_{\B}(\tau) \bigr]^{ }_\rmi{renorm,$\bmu$}
 \;, \la{def_GB_physical} \\[2mm] 
 c^{ }_{\B}(\bmu)  & = &
 1 + \frac{ g^2 \CA }{8\pi^2} 
 \biggl[ 
     \ln\biggl( \frac{\bmu e^{\gammaE}}{4\pi T} \biggr)
   - 1 
 \biggr] 
 + \rmO(g^4) 
 \;. \la{cB}
\ea
The information in \eq\nr{cB} can be rephrased by running 
$\msbar$ operators to the scale~\cite{Bmatch}
\be 
 \bmu \approx 19.179 T
 \;, \la{thermal_scale} 
\ee
where the square bracket in \eq\nr{cB} vanishes. We note that this 
is a remarkably large scale, suggesting a reasonable convergence 
down to fairly low temperatures. 

After the establishment of \eq\nr{thermal_scale}, the remaining 
challenge is to convert the lattice operator in \eq\nr{B_bare} into
the $\msbar$ scheme. This problem has been 
addressed in ref.~\cite{renormB}, in the context of 
a spin-dependent operator involving the magnetic field~\cite{hqet1}. 
The procedure involves the use of a finite-volume scheme with 
Schr\"odinger functional (SF) boundary conditions; 
the conversion of those results into a renormalization group 
invariant (RGI) operator; and a subsequent relation of the RGI operator
to the $\msbar$ one at the scale $\bmu$. 
 The conversions are achieved by multiplying the physical correlation
 function by ratios of unphysical but technically more accessible 
 auxiliary correlation functions, denoted below by $\Phi$, which
 can be determined in different renormalization schemes.
The ingredients are described in more detail in appendix~A. 
The end result can be expressed as  
\be
\frac{ [ G^{ }_{\B}(\tau) ]^{ }_\rmi{physical} }
     { [ G^{ }_{\B}(\tau) ]^{ }_{\rmi{bare},\rmii{L}} }
 = 
 \biggl\{ 
    \frac{\Phi^{ }_\rmii{$\msbar$}(\bmu = 19.179 T)}
         {\Phi^{ }_\rmii{RGI}}
  \times  
 \frac{\Phi^{ }_\rmii{RGI}}{\Phi^{ }_\rmii{SF}(\frac{1}{2 L^{ }_\rmi{max}})}
  \times
  Z^\rmii{SF}_\rmi{spin} ( 2 L^{ }_\rmi{max} )
 \biggr\}^2_{ } 
 \;, \la{master}
\ee
where the separate factors originate from
\eqs\nr{eq57} and \nr{eq67}.

We end this section by noting that, 
numerically, the renormalization 
factor in \eq\nr{master} is fairly large.
Typical values entering our analysis are
$\{...\}^2 \approx 3.12$ for 
$\beta \approx 7.0$ and $T \approx 1.2\Tc$, or 
$\{...\}^2 \approx 2.87$ for 
$\beta \approx 7.6$ and $T \approx 2.0\Tc$.

%
\section{Numerical implementation}
\la{se:methods}

%
\begin{table}[t]

\small{
\begin{center}
\begin{tabular}{cccccccccc} 
 $\beta$ & 
 $\Nt$ &  $\Ns$ &
 subs & 
 sub-ups & 
 confs & 
 streams & 
 $\tau^{ }_\rmi{int}$ &  
 $ \rO / a $ & 
 $ T / \Tc  $ 
 \\[3mm]
 \hline 
  6.860  & 20 & 48 & 
  5 & 
  500 & 
  500 & 
  5  & 
  13  & 
  17.7  &
  1.19
  \\[-1mm] 
     & 20 & 56 & 
  5 & 
  500 & 
  905 & 
  5  & 
  13  & 
  17.7  &
  1.19
  \\[-1mm] 
     & 20 & 64 & 
  5  & 
  500  & 
  1020 & 
  6  & 
  11  & 
  17.7    &
  1.19
  \\
  7.010 
         & 24 &  64 &
  6 & 
  500 & 
  1465 & 
  10  & 
  7  &
  21.3 &
  1.19  \\[-1mm] 
         &  24 &  72 &
  6  & 
  500  & 
  1051 & 
  15  & 
  4  &
  21.3 &
  1.19  \\
  7.050  & 20 &  48 &
  5 & 
  500 & 
  875 & 
  4  & 
  5  &
  22.4   &
  1.50   \\[-1mm]
       & 20 &  56 &
  5 & 
  500 & 
  706 & 
  5  & 
  5  &
  22.4   &
  1.50   \\[-1mm]
        & 20 &  64 &
  5 & 
  500 & 
  1000 & 
  8  & 
  8  &
  22.4   &
  1.50   \\
  7.135  & 28 &  84 &
  7 & 
  1000 & 
  1225 & 
  9  & 
  11  &
  24.8   &
  1.19   \\
  7.192 
     & 24 &  60 & 
         4 & 
        500 & 
        1530 & 
        9    & 
        5    &
        26.6   &
        1.48  \\[-1mm] 
    & 24  &  72 & 
  4  & 
  500  & 
  1448 & 
  7   & 
  5   &
  26.6   &
  1.48  \\[-1mm] 
    & 30 &  96 & 
  5 & 
  1000 & 
  1256 & 
  12 & 
  12 &
  26.6 &
  1.19  \\
  7.300  &  20 &  48 &
  4 & 
  500 & 
  1000 & 
  4   & 
  5   &  
  30.2   &
  2.03  \\[-1mm]
         &  20 &  64 &
  4 & 
  500 & 
  1120 & 
  8 & 
  5  &  
  30.2   &
  2.03  \\
  7.330  &  28 &  84 &
  7 & 
  1000 & 
  1256 & 
  10   & 
  11   &  
  31.3   &
  1.50  \\
  7.457  &  24 &  60 &
  4 & 
  500 & 
  1645 & 
  9 & 
  4  &  
  36.4   &
  2.04  \\[-1mm]
    &  24 &  72 &
  4 & 
  500 & 
  1038 & 
  7 & 
  4  &  
  36.4   &
  2.04  \\
  7.634  & 30 &  96 & 
  5 & 
  1000 & 
  1130 & 
  9 & 
  7 &
  44.9 &
  2.01  \\ 
 \hline 
\end{tabular} 
\end{center}
}

\vspace*{3mm}

\caption[a]{\small
  The lattices simulated
  (coupling $\beta = 6/g_0^2$, geometry $\Nt\times\Ns^3$),
  together with  
  sublattices (subs), 
  sublattice updates (sub-ups), 
  statistics (confs), 
  streams, 
  integrated autocorrelation times
  ($\tau^{ }_\rmi{int}$), 
  and physical scales 
  ($\rO/a$~\cite{rs}, $T/\Tc^{ }$~\cite{Tc}). 
  The last two are estimated
  as explained below \eq\nr{scale_setting}. 
  The measurements obtained from all of these lattices 
  are attached to this submission as ancillary files. 
 }
\label{table:params}
\end{table}
%

We now proceed with the numerical measurement of the correlation 
function defined by \eqs\nr{GB_def}--\nr{plaquette}. The measurements
are carried out in quenched QCD, with the Wilson gauge action, parametrized
by a bare coupling $\beta = 6 / g_\rmi{0}^2$. The $\beta$ values and  
lattice volumes are assembled 
in table~\ref{table:params}. We note that spectral studies 
necessitate a large number of temporal points and 
correspondingly fine lattices (large values of $\beta$). 
As a consequence our physical volumes are small compared with the 
state-of-the-art~\cite{mesonic}, nevertheless the finite-volume effects 
appear to be somewhat smaller than statistical uncertainties (see below). 

For a conversion to physical units~\cite{rs}, 
we have employed the interpolating function 
\be
 \ln\Bigl( \frac{\rO}{a} \Bigr)
 = 
 \biggl[\frac{\beta}{12 b^{ }_0} + \frac{b^{ }_1}{2 b_0^2}
  \ln\Bigl( \frac{6 b^{ }_0}{\beta} \Bigr)  \biggr]
 \frac{1 + c^{ }_1 /\beta + c^{ }_2 / \beta^2}
      {1 + c^{ }_3/\beta + c^{ }_4 / \beta^2}
 \;, \quad
 b^{ }_0 \equiv \frac{11}{(4\pi)^2}
 \;, \quad 
 b^{ }_1 \equiv \frac{102}{(4\pi)^4}
 \;, \la{scale_setting}
\ee
inserting updated coefficients for large $\beta$
from the caption of table~2 of ref.~\cite{GPtau}, {\it viz.}\ 
\be
    c^{ }_1 = -8.9664
 \;, \quad
    c^{ }_2 = 19.21
 \;, \quad
    c^{ }_3 = -5.25217 
 \;, \quad
    c^{ }_4 = 0.606828
 \;. 
\ee
Conversions to units 
of $\Tc$ make use of 
$\rO \Tc = 0.7457(45)$ 
from ref.~\cite{Tc}.

As far as the Monte Carlo update goes, 
the lattice was divided into sublattices along the~$\tau$~direction, 
in accordance with the multilevel philosophy~\cite{lw,shear}. 
Additional sublattice updates were performed.
Each update consisted of 1~heatbath step followed
by 3~overrelaxation steps. We started a number of streams 
for each parameter set, with different pseudorandom number chains. 
The runs were started from a cold configuration, in order to give
the Polyakov loop a real expectation value and to stay in 
the sector of trivial topology. 

To probe the efficiency of the update, we calculated 
an integrated autocorrelation time, $\tau^{ }_\rmi{int}$, for the
absolute value of the Polyakov loop. Data from different streams 
were combined according to ref.~\cite{wolff}. The integration 
extended to a time interval in which the autocorrelation function
had not decreased below $10^{-3}$ of its original value. 
Sufficient thermalization 
was allowed for ($ \gg 100 \tau^{ }_\rmi{int}$). 
The parameters of the runs are listed in table~\ref{table:params}.

We also looked at autocorrelations in the numerator of 
$
 G^{ }_{\B}(\tau) 
$ (cf.\ \eq\nr{GB_def}). 
For small~$\tau$, we got autocorrelation
times similar to those from the Polyakov loop. For larger~$\tau$, 
the measurement of the autocorrelation function became noisy, however
as long as a signal could be obtained, the autocorrelation time 
decreased rather than increased with $\tau$. 

A much longer autocorrelation time is known to characterize
the topological charge, $Q$~\cite{slow}. 
We checked its movement on some of our lattices, 
defining $Q$ via gradient flow~\cite{t0}. 
Our lattices are at high temperatures, where the physical topological 
susceptibility becomes vanishingly small. Starting from 
a configuration with all links set to unity, we found that $Q$
sticks to the trivial sector.  
We also checked that the final configurations for these sets are at $Q=0$. 
This suggests that our configurations were stuck in the
trivial sector all along.

The values of $\tau^{ }_\rmi{int}$ as indicated 
in table~\ref{table:params}
were taken as estimators of the autocorrelations in the data. 
Subsequently the data was blocked in 
block sizes $ \gsim 2 \tau^{ }_\rmi{int}$, 
and the blocked data was considered independent 
in the bootstrap analysis. 

%
\section{Data and its analysis}
\la{se:data}

The methods and statistics described in the previous section lead to 
a good signal for the observable in \eq\nr{GB_def}, 
in the temperature range indicated in table~\ref{table:params}. 
In this section we explain how we have estimated 
the infinite-volume and continuum limits of this data set. 

\begin{figure}[t]

\hspace*{-0.1cm}
\centerline{%
    \epsfxsize=5.0cm\epsfbox{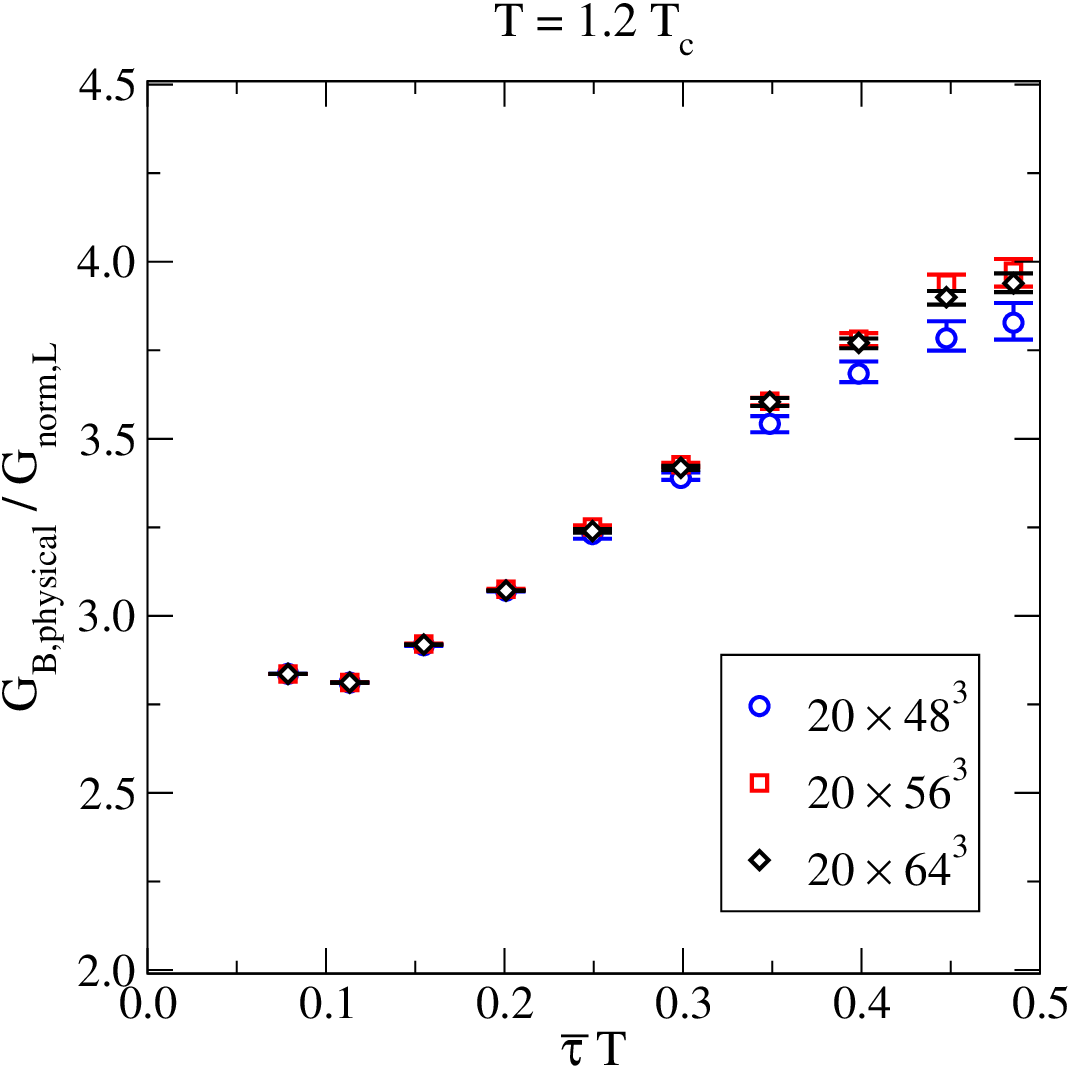}
  ~~\epsfxsize=5.0cm\epsfbox{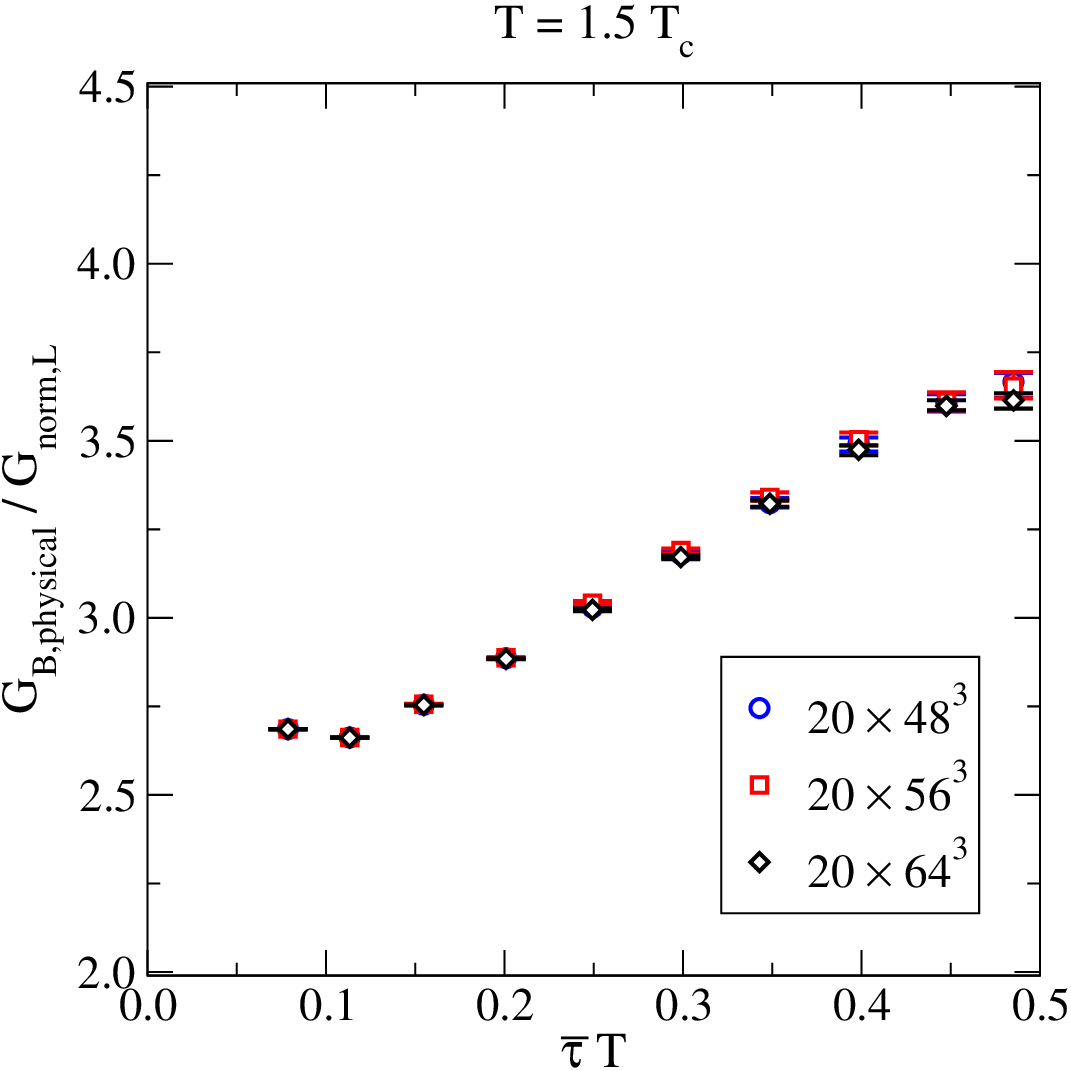}
  ~~\epsfxsize=5.0cm\epsfbox{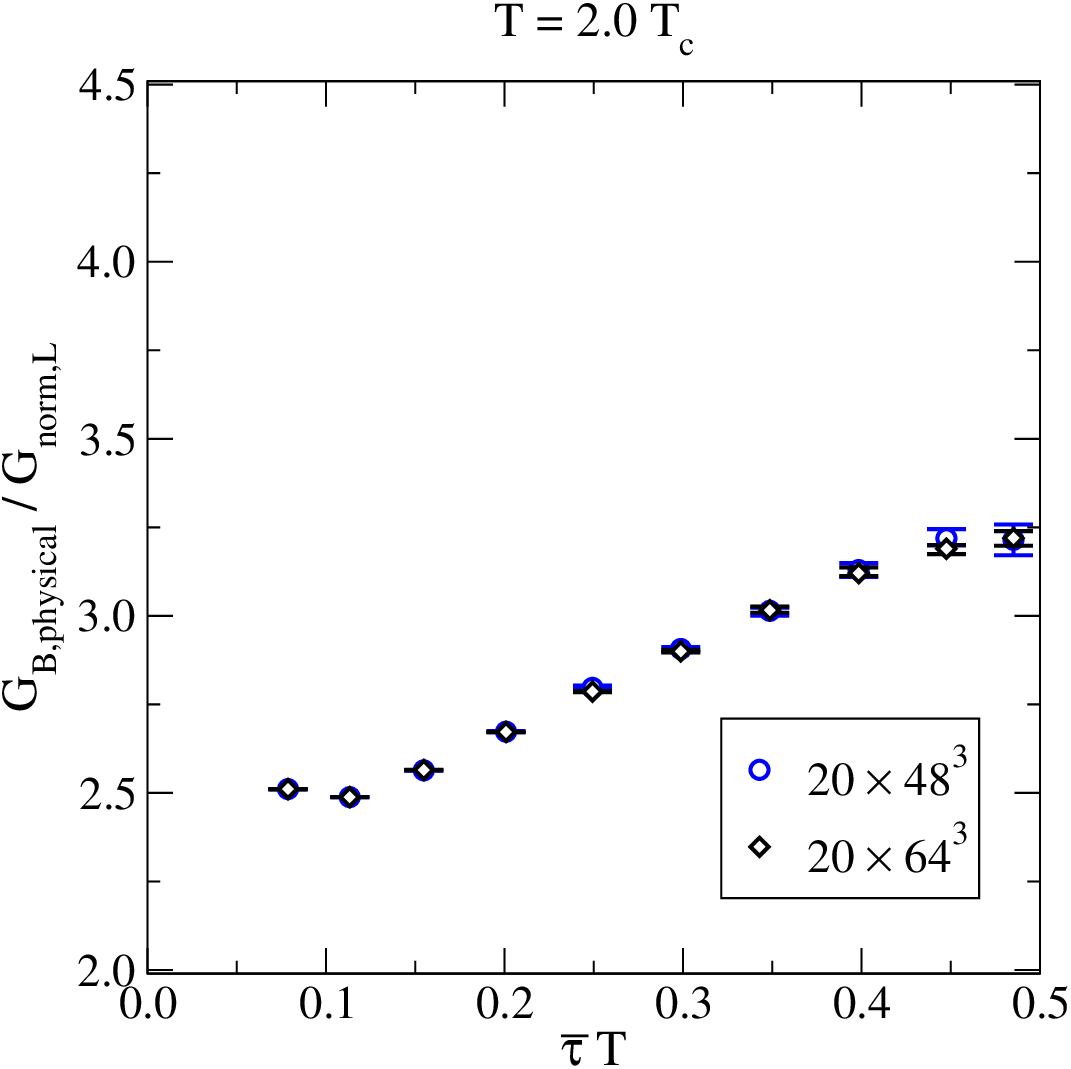}
}

\caption[a]{\small
     Illustration of the volume dependence of the 
     renormalized correlator (cf.\ \eq\nr{master}) at 
     $T\approx 1.2\Tc^{ }$ (left),
     $1.5\Tc^{ }$ (middle), and
     $2.0\Tc^{ }$ (right),
     after normalization to \eq\nr{G_norm_L}.
     By $\overline{\tau}$ we denote an improved distance, 
     as defined in \eq\nr{improvement}.
}

\la{fig:volume}
\end{figure}

In order to show the results of the measurements, 
we normalize them to the correlator obtained with 
tree-level lattice perturbation theory, 
cf.\ \eq\nr{G_norm_L}. 
Results from different volumes are shown in \fig\ref{fig:volume}. 
We observe no appreciable 
evolution at $\Ns \gsim 3 \Nt$, and conclude that such
results reasonably approximate the infinite-volume limit 
within statistical uncertainties, 
even if our volumes are admittedly small in physical units. 

\begin{figure}[t]

\hspace*{-0.1cm}
\centerline{%
   \epsfxsize=5.0cm\epsfbox{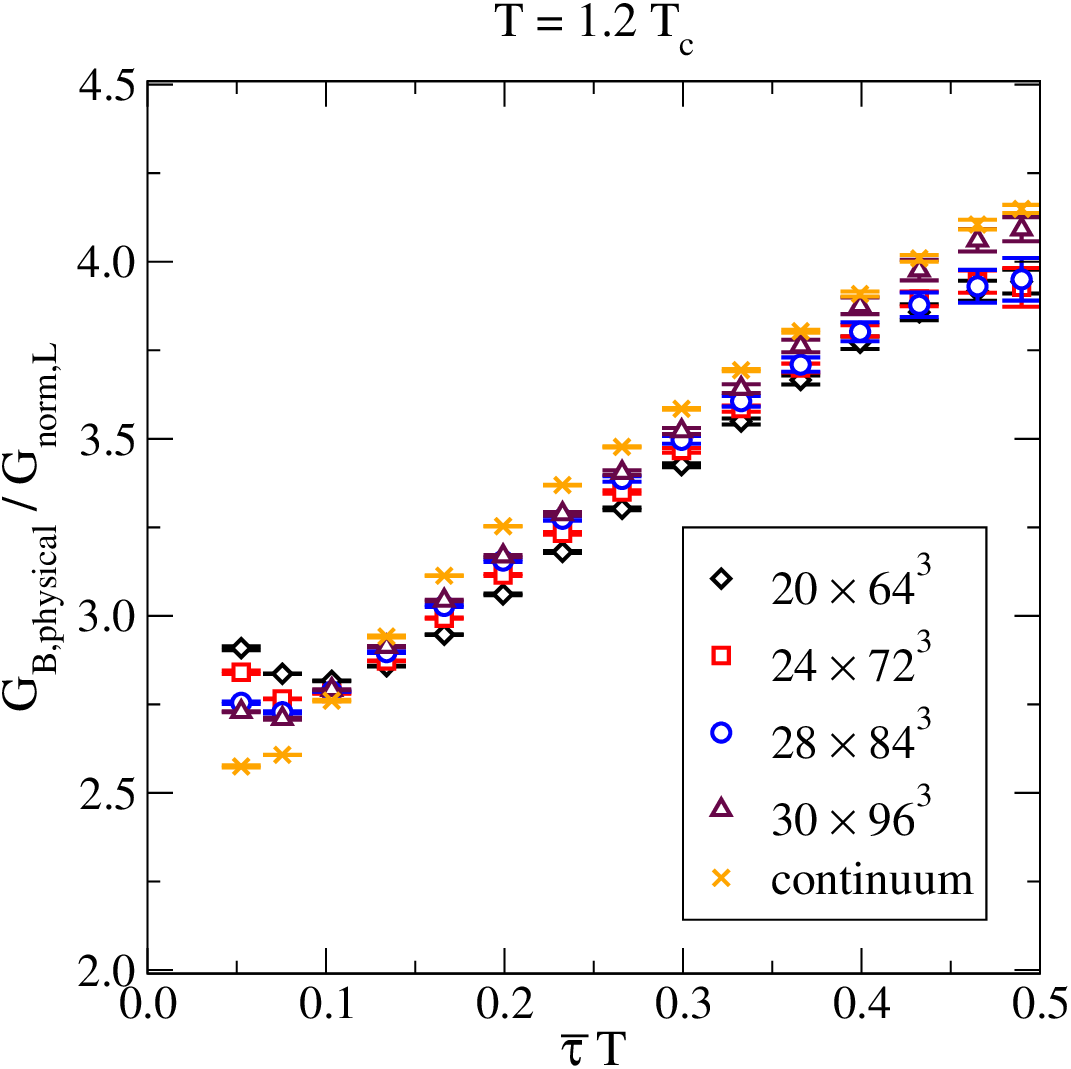}
 ~~\epsfxsize=5.0cm\epsfbox{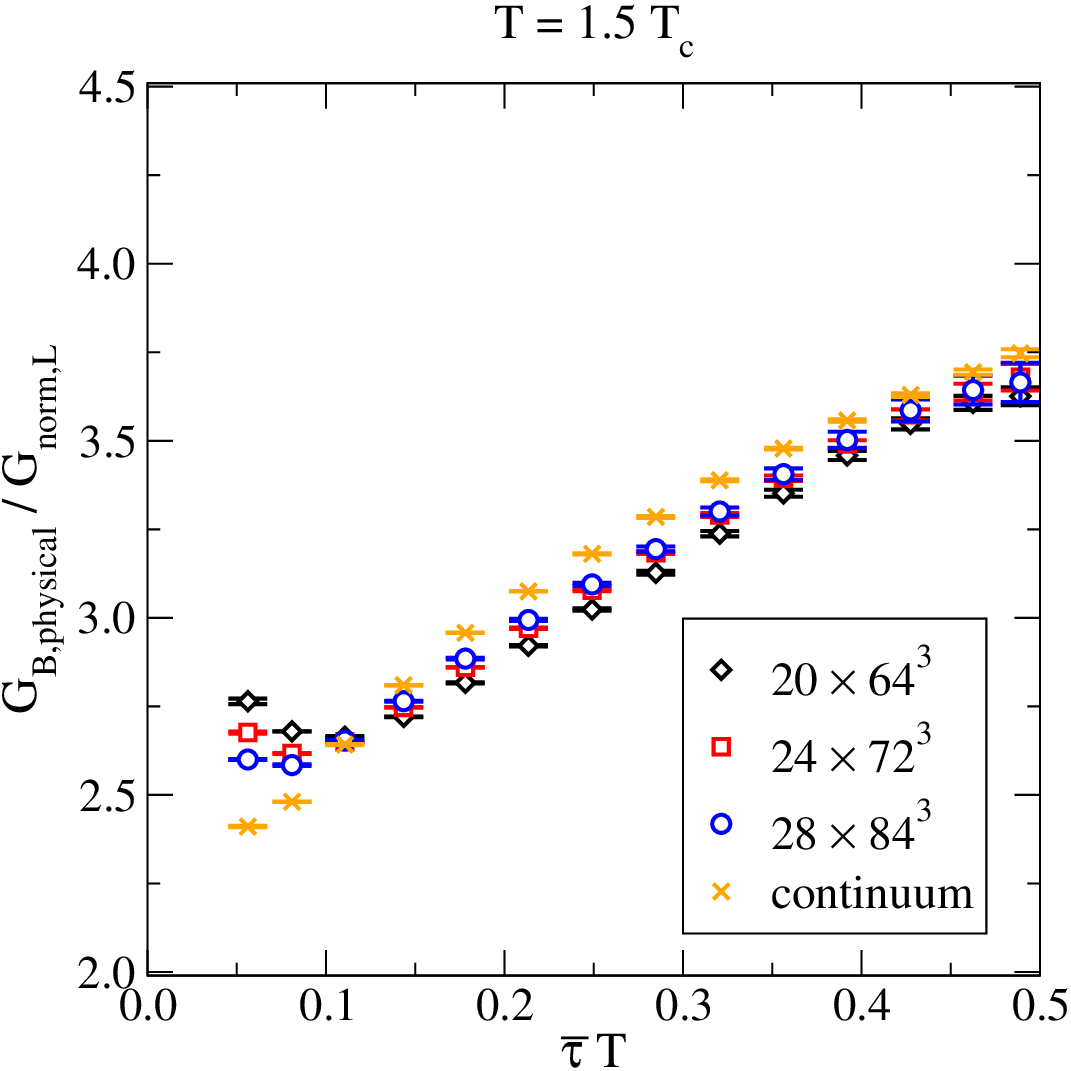}
 ~~\epsfxsize=5.0cm\epsfbox{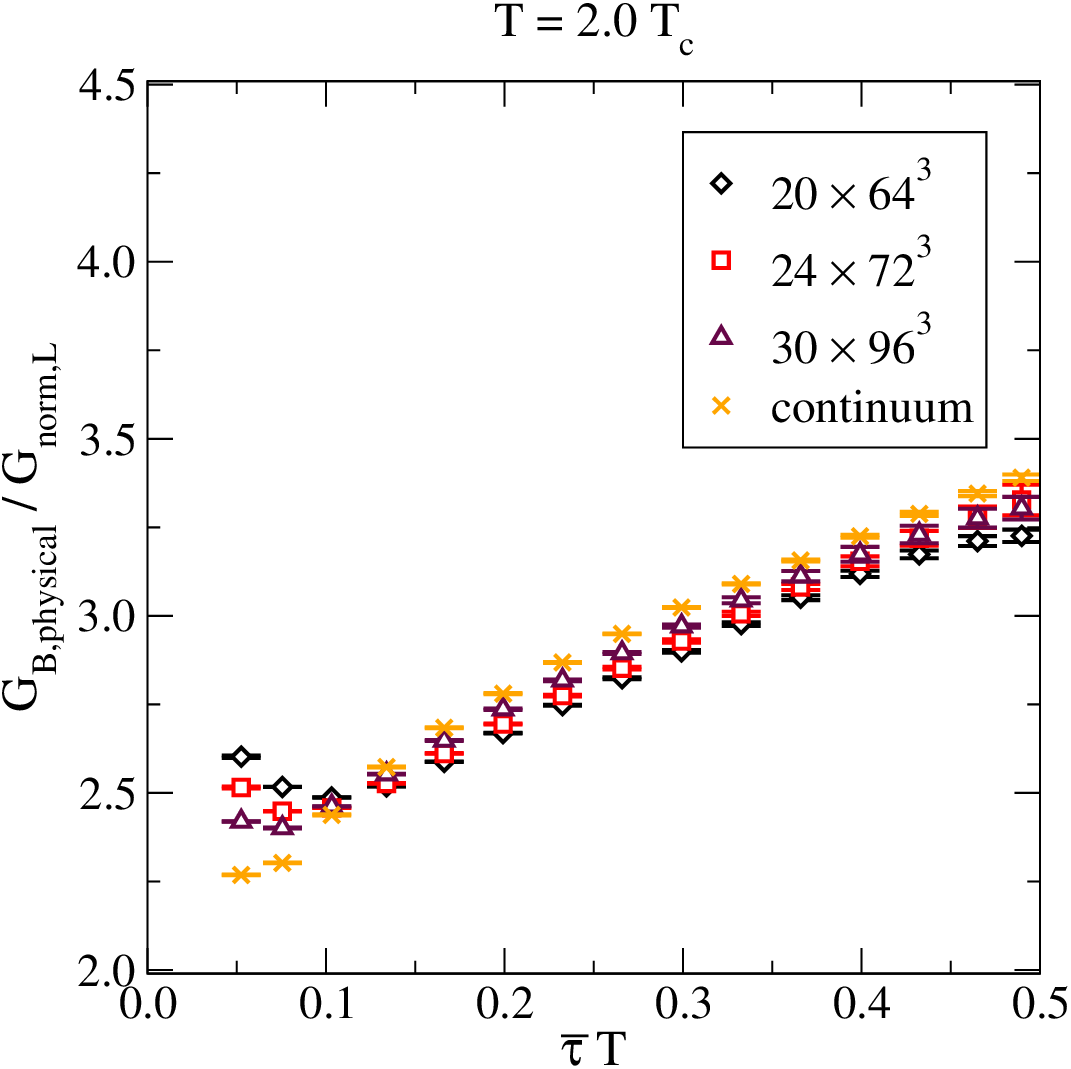}
}

\caption[a]{\small
     Illustration of the approach of the renormalized 
     correlator (cf.\ \eq\nr{master}) to the continuum limit, 
     at $T\simeq 1.2\Tc^{ }$ (left),  
     $1.5\Tc^{ }$ (middle), 
     and $2.0\Tc^{ }$ (right), 
     after normalization to \eq\nr{G_norm_L}. 
     Data at the coarser lattices have been 
     interpolated to the distances of the finest lattice. 
}

\la{fig:continuum}
\end{figure}

In order to extrapolate to continuum, the bare data need to be 
multiplied with the renormalization factor from \eq\nr{master}.  
In \fig\ref{fig:continuum}, 
we show results from different lattice spacings
(at more or less fixed aspect ratios).
Only modest 
lattice spacing dependence can be discerned within 
the statistical uncertainties. 

To accelerate the approach 
to the continuum limit, we have assigned the bare lattice
data to tree-level improved distances~\cite{rs,hbm} (cf.\ appendix~B), 
denoted by $\overline{\tau}$ in \figs\ref{fig:volume}
and~\ref{fig:continuum}.
The data at the coarser lattices are B-spline interpolated (or in
single cases, extrapolated) 
to the improved distances of the finest lattice. 
Subsequently fits linear 
in $1/N^{2}_\tau$ are carried out. 
The extrapolation results
are shown in \fig\ref{fig:continuum} with the crosses.

Finally, 
the ratio obtained is multiplied by the continuum version 
of \eq\nr{G_norm_L}, {\it viz.}\  
\be
 G^\rmii{ }_{\rmi{norm},\rmii{DR}} (\tau)
 \; \equiv \; 
 \pi^2 T^4 \left[
 \frac{\cos^2(\pi \tau T)}{\sin^4(\pi \tau T)}
 +\frac{1}{3\sin^2(\pi \tau T)} \right] 
 \;. \la{G_norm_DR}
\ee
Thereby we obtain an estimate of the continuum colour-magnetic 
correlator that can be subjected to a spectral analysis. 

%
\section{Implications for heavy quark momentum diffusion}
\la{se:implications}

Having estimated the continuum limit of $G^{ }_{\B}$ in the previous
section, the remaining challenge is to extract the momentum diffusion
coefficient from this data. If the continuum correlator is expressed
in a spectral representation, 
\be
 \bigl[ G^{ }_{\B}(\tau) \bigr]^{ }_\rmi{physical} \; \equiv \; 
 \int_0^\infty
 \frac{{\rm d}\omega}{\pi} \rho^{ }_{\B} (\omega)
 \frac{\cosh \bigl[ \bigl(\frac{1}{2T} - \tau\bigr)\omega \bigr] }
 {\sinh\left( \frac{\omega}{2 T} \right) } 
 \;, \la{rhoB_def}
\ee
then $\kappa^{ }_{\B}$ is given by 
\be
 \kappa^{ }_{\B} = \lim_{\omega\to 0} 
 \frac{2T\rho^{ }_{\B}(\omega)}{\omega}
 \;. \la{kappaB_def}
\ee

The procedure that we adopt for extracting $\kappa^{ }_{\B}$ goes through
modelling the shape of $\rho^{ }_{\B}$. Similarly to the 
colour-electric spectral function~\cite{mink}, we do not expect 
$\rho^{ }_{\B}$ to contain any sharp transport peak at $\omega \ll T$.
Then, its infrared (IR) part can be approximated as  
\be
   \phi^{ }_\rmii{IR}(\omega) \; \equiv \; 
   \frac{\kappa^{ }_{\B}\, \omega}{2 T}
 \;. \la{phiIR}   
\ee
For the ultraviolet (UV) part, we adopt the vacuum-like ansatz
\be
  \phi^{ }_\rmii{UV}(\omega) \; \equiv \;
  \frac{g^2(\bmu^{ }_{\B})\, \CF\, \omega^3}{6\pi} 
 \;, \quad
 \bmu^{ }_{\B} \equiv 
 \mbox{max}\Bigl[ 
    \omega^{1 - \frac{\gamma_0}{b_0}}_{} (\pi T)^{\frac{\gamma_0}{b_0}}
   ,\pi T \Bigr]
 \;, \la{phiUV}
\ee
where 
$
 \CF \equiv ({\Nc^2 - 1})/({2 \Nc}) 
$ 
and, 
for $\{ \Nc^{ },\Nf^{ } \} = \{ 3,0 \}$, 
the colour-magnetic anomalous dimension is 
$\gamma^{ }_0 \equiv 3 / (8\pi^2)$  (cf.\ appendix~A), 
and $b^{ }_0 = 11/(16 \pi^2)$.\footnote{%
  More precisely, the spectral function has the shape
  $
   \rho^{ }_{\iB}(\omega)
   =
   \frac{ g^2(\bmu) C_\rmiii{F}^{ }\, \omega^3 }{6\pi}
   \bigl\{
     1 + g^2(\bmu)
       \bigl[
         (b^{ }_0 - \gamma^{ }_0) \ln\frac{\bmu^2}{\omega^2}
	  + \gamma^{ }_0 \ln\frac{\bmu^2}{(\pi T)^2 }
	  + c + f^{ }_\T \bigl( \frac{\omega}{\pi T} \bigr)
       \bigr]
   \bigr\} 
  $, 
  cf.\ \eqs\nr{def_GB_physical} and \nr{rhoB_nlo}, 
  where $f^{ }_\T$ is power-suppressed at
  large values of $\omega/(\pi T)$.
  Eq.~\nr{phiUV} follows by choosing
  the renormalization scale $\bmu$
  so as to eliminate large logarithms. 
 }  
In the colour-electric case, $\gamma^{ }_0$ is absent~\cite{rhoE}, 
whereby $\bmu^{ }_{\E} \equiv \mbox{max}(\omega,\pi T)$.
The model spectral function is defined as 
\be
 \rho^{ }_{\B}(\omega) \; \equiv \; 
 \sqrt{
    {\phi}^{2}_\rmii{IR}(\omega) \; + \; 
    a^{ }_{\B}\, \phi^{2}_\rmii{UV}(\omega) 
 }
 \;, \la{model_B}
\ee
where $a^{ }_{\B} \simeq 1$ is treated as a fit parameter. 
This type of a model, 
together with many other shapes,\footnote{%
 We have tested some of them, remaining with two fit parameters, 
 and included the spread in our results.  
 The tests are described in more detail in appendix~D.
 } 
were explored for the 
colour-electric case in ref.~\cite{lat4}. The results 
were found to lie around the middle of 
an admissible range, even if the systematic uncertainties 
are underestimated by the 
very constrained ansatz of \eq\nr{model_B}. 

\begin{figure}[t]

\hspace*{-0.1cm}
\centerline{%
   \epsfxsize=7.5cm\epsfbox{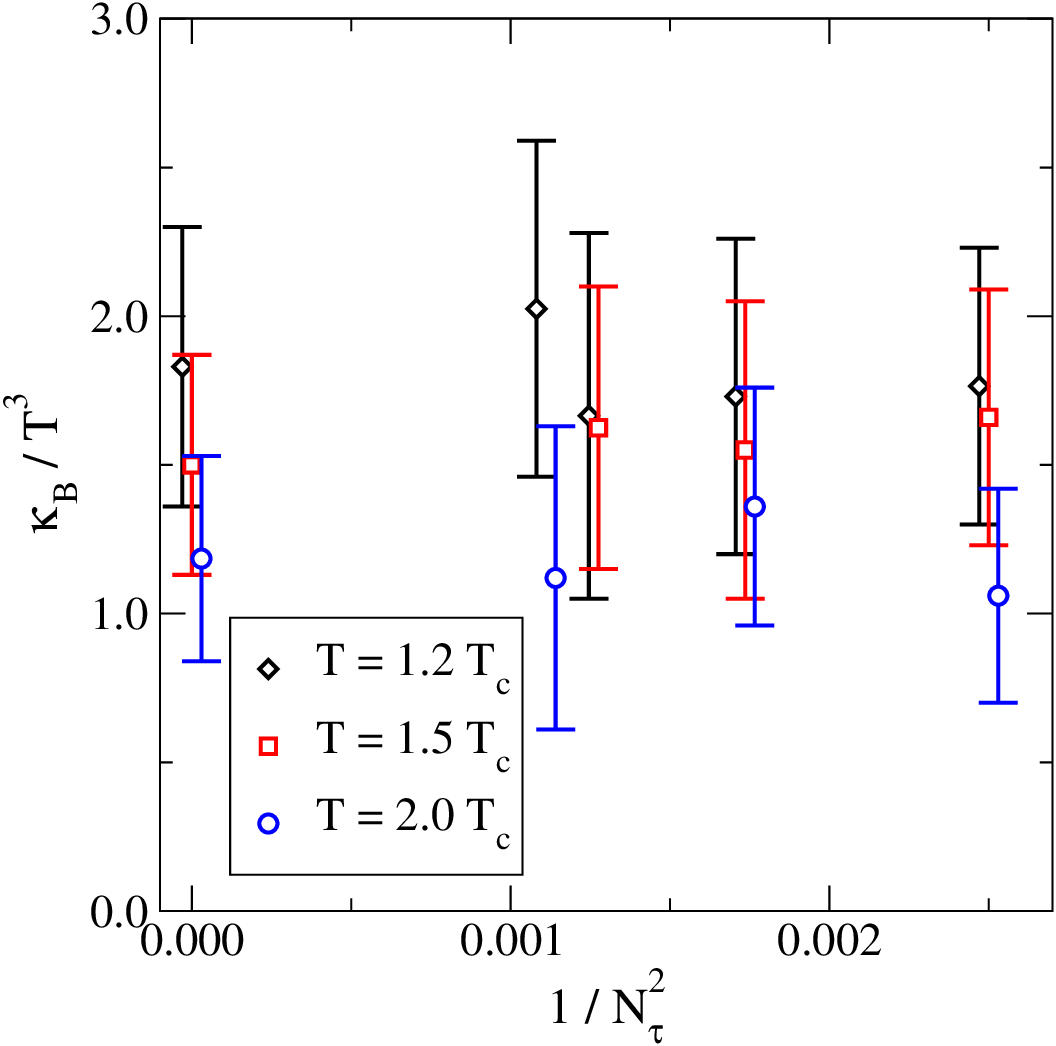}
}

\caption[a]{\small
     The parameter $\kappa^{ }_{\B} / T^3$ as obtained from
     finite-$a$ correlators, and separately from
     the continuum-extrapolated correlator. For better visibility, 
     the data have been slightly displaced in the horizontal direction. 
     Discretization uncertainties are modest compared with statistical ones. 
}

\la{fig:results}
\end{figure}

Inserting \eq\nr{model_B} into \eq\nr{rhoB_def}, 
$G^{ }_{\B}(\tau)$ is a function of two 
parameters, $\kappa^{ }_{\B}$ and $a^{ }_{\B}$. 
We fit the result to both finite-$a$ and to 
the continuum-extrapolated data sets 
in the range $\tau \ge \tau^{ }_\rmi{min} \in (0.20 - 0.35)/T$, 
with $ \tau^{ }_\rmi{min} $
chosen so as to obtain good $\chi^2/$d.o.f.
Error estimates are based on a bootstrap analysis on
the blocked data (cf.\ \se\ref{se:methods}), 
with correlations in the data at different values of $\tau$ 
accounted for in the finite-$a$ case, 
as well as on differences originating from 
variations in $\tau^{ }_\rmi{min}$.
Results are illustrated in \fig\ref{fig:results}, whose
error bands encompass other fit forms as well,
as described in appendix~D.

\begin{figure}[t]

\hspace*{-0.1cm}
\centerline{%
   \epsfysize=7.5cm\epsfbox{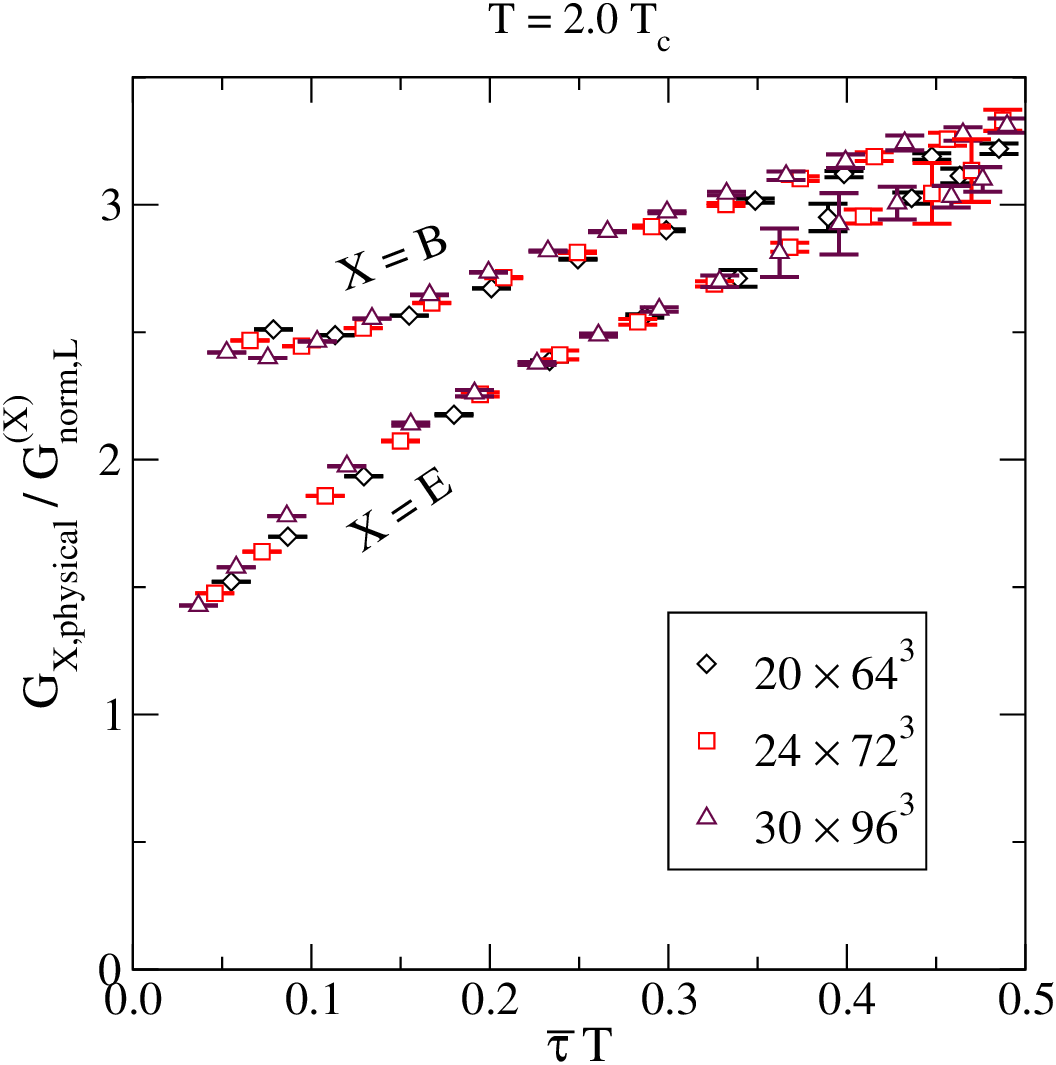}
 ~~\epsfysize=7.5cm\epsfbox{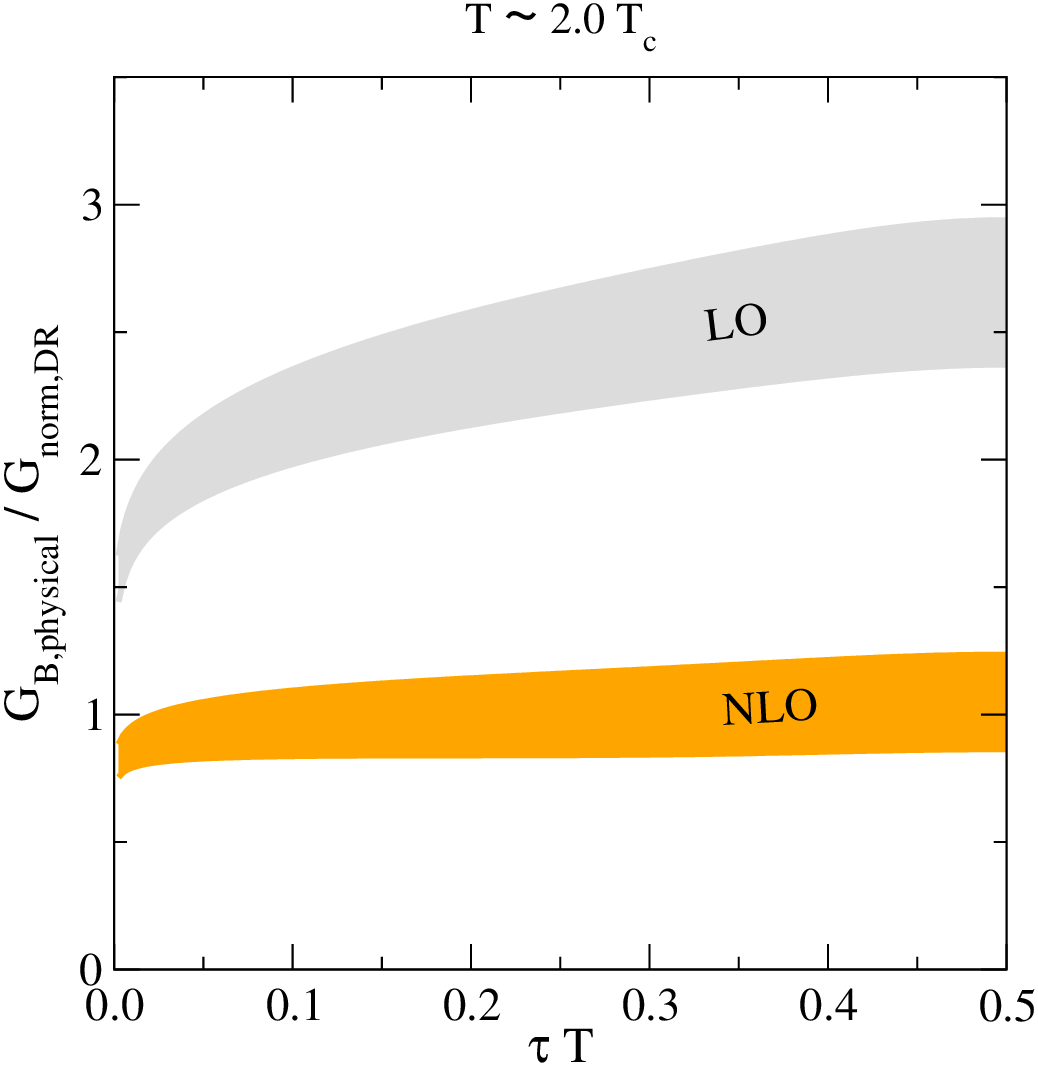}
}

\caption[a]{\small
     Left: comparison of the colour-magnetic
     and colour-electric correlators, at $T = 2.0\Tc^{ }$
     (improved distances and normalizations 
     are different in the two cases, cf.\ appendix~B). 
     Right: the colour-magnetic correlator according
     to NLO perturbation theory, cf.\ appendix~C.
     The lattice data 
     suggests a large contribution on top of the NLO result
     (the latter shows poor convergence, which could be 
     marginally improved upon by tuning the numerical factors
     in the scale choice in \eq\nr{phiUV}). 
}

\la{fig:EE}
\end{figure}

We have repeated the analysis for the 
colour-electric correlator with pre-existing data
and with a part of our new set, 
permitting for a direct comparison of the two 
transport coefficients. In this case, renormalization
is perturbative~\cite{renormE},
 as non-perturbative renormalization factors have
 unfortunately not been worked out to date. The correlation
functions are illustrated in \fig\ref{fig:EE}, which also shows
a comparison with NLO perturbative results from appendix~C.

%
\begin{table}[t]

\small{
\begin{center}
\begin{tabular}{ccccccccc} 
 $\!\! T / \Tc^{ }  \!\!\!$ & 
 $\kappa^{ }_{\E}/T^3$ &  
 $\kappa^{ }_{\B}/T^3$ & 
 $\!\! \langle\vec{v}^2 \rangle^{ }_\rmi{c} \!\!$ & 
 $\!\! \langle\vec{v}^2 \rangle^{ }_\rmi{b} \!\!$ & 
 $\kappa^{ }_\rmi{c}/T^3$ & 
 $\kappa^{ }_\rmi{b}/T^3$ &
 $\eta^{ }_\rmi{c}/T$ & 
 $\eta^{ }_\rmi{b}/T$ 
 \\[2mm]
 \hline
 \\[-4mm] 
  1.2 & 
  ${1.5} - {3.4}^*_{ }$   & 
  $1.0 - 2.6$ & 
  0.52  & 
  0.20  & 
  $1.8 - 4.3$  &  
  $1.6 - 3.8$  &  
  $0.16 - 0.38$  & 
  $0.05 - 0.13$ \\[0.5mm]  
  1.5 & 
  ${1.3} - {2.8}^*_{ }$  & 
  $1.0 - 2.1$ & 
  0.59 & 
  0.24 & 
  $1.7 - 3.7$ & 
  $1.4 - 3.2$ & 
  $0.16 - 0.36$  & 
  $0.05 - 0.13$ \\[0.5mm]  
  2.0 & 
  ${1.0} - {2.5}^*_{ }$  & 
  $0.6 - 1.8$ & 
  0.67  & 
  0.30  & 
  $1.2 - 3.4$ & 
  $1.1 - 2.9$ & 
  $0.14 - 0.37$  & 
  $0.05 - 0.15$ \\  
 \hline 
\end{tabular} 
\end{center}
}


\caption[a]{\small
 Fit results for the contributions to $\kappa$ from 
 the colour-electric and colour-magnetic fields. 
 Results for $\kappa^{ }_{\E}$ have been indicated with a star 
 because measurements were only carried out on a subset of our lattices
 and thus some finite-volume and finite-$a$ checks are missing.
 For comparison, a more comprehensive analysis suggests 
 $\kappa^{ }_{\E}/T^3 \simeq 1.8 - 3.4 $ 
 at $T/\Tc^{ } = 1.5$~\cite{lat4}.  
 In order to combine the results according to \eq\nr{kappa_full}, 
 the average thermal velocity is needed; this has been 
 estimated from NLO results for the constant parts 
 in the spatial and temporal vector current correlators~\cite{GVtau}.
 The drag force~$\eta$ is  
 obtained from \eq\nr{eta}. 
 The subscripts c,b refer to charm and bottom quarks.  
 }
\label{table:results}
\end{table}
%

Our final results for the transport coefficients are collected
in table~\ref{table:results}
(the errors shown include the full spreads from \fig\ref{fig:results}). 
The electric and magnetic contributions
have been combined according to~\cite{1overM}
\be
 \kappa^{ }_{ } \;\simeq\; 
 \kappa^{ }_{\E} + 
 \frac{2}{3} \langle \vec{v}^{2}_{ } \rangle \, 
 \kappa^{ }_{\B}
 \;, \la{kappa_full}
\ee
which requires knowledge of the average thermal velocity. 
Within the $T/M$ expansion, this can be expressed as~\cite{1overM}
\be
 \langle \vec{v}^2 \rangle 
 \approx \frac{3T}{M_\rmi{kin}}
 \biggl(  
  1 - \frac{ 5 T }{ 2 M_\rmi{kin} }
 \biggr)
 \;, 
\ee
however the value of the kinetic mass $M^{ }_\rmi{kin}$ is ambiguous, 
and the large coefficient in the round brackets renders
the convergence of the expansion questionable. An alternative
definition of~$\langle \vec{v}^2 \rangle$ can be given 
as the ratio of the almost constant ($\tau$-independent) 
part of the spatial vector current correlator, 
and the susceptibility~\cite{kappaE}. 
This definition
does not rely on an expansion in $T/M$, and has been worked out up to 
NLO in the weak-coupling expansion~\cite{GVtau}. The corresponding
numerical values have been indicated in table~\ref{table:results}.\footnote{%
  Separate values for the numerator (i.e.\ spatial vector correlator)
  and denominator (i.e.\ susceptibility) 
  can be found on the web page
  {\tt http://www.laine.itp.unibe.ch/quarkonium/}
}
 
Given the value of $\kappa$, the drag force $\eta$ appearing in 
\eq\nr{langevin} can be obtained from a fluctuation-dissipation relation.   
A convenient way to express this, accurate up to NLO corrections 
in $T/M^{ }_\rmi{kin}$~\cite{1overM}, reads
\be
 \eta 
 \approx
 \frac{\kappa^{ }_{ }\, \langle\vec{v}^2\rangle}{6 T^2}
 \;. \la{eta}
\ee
This has also been shown in table~\ref{table:results}.
The traditional diffusion coefficient could similarly 
be estimated from $D \simeq 2 T^2/\kappa^{ }_{ }$, 
however it plays no direct role in studies based on \eq\nr{langevin}.

%
\section{Conclusions and outlook}
\la{se:concl}

The purpose of this paper has been to present a numerical simulation
of the colour-magnetic correlator defined by \eq\nr{GB_def}, followed
by its non-perturbative renormalization (cf.\ \eq\nr{master}) and 
extrapolation to the continuum limit (cf.\ \se\ref{se:data}). 
Subjecting the result to simple-minded spectral modelling 
(cf.\ \se\ref{se:implications}), yields physical 
results as displayed in table~\ref{table:results}.

A strength of our effective-theory approach, 
based on an expansion in $T/M$, 
is that the spectral function corresponding to \eq\nr{GB_def}
is believed to be smooth at small $\omega \ll T$, i.e.\ free from a sharp
transport peak. If the corresponding physics is addressed with
a relativistic formulation instead, measuring vector current correlators, 
the spectral function has a transport peak of width 
$\eta \sim \alphas^2 T^2 / M$~\cite{pt}. It has been suggested recently 
that employing a Lorentzian-shaped spectral function, 
of width $\eta$, in connection with lattice data, meaningful
constraints on $\eta$ can be extracted~\cite{D_relativistic}.
Specifically, table VI of ref.~\cite{D_relativistic} cites 
$\eta^{ }_\rmi{c}/T \simeq 0.9 - 5.3$ at $T/\Tc^{ } = 1.5 - 2.25$, 
and 
$\eta^{ }_\rmi{b}/T \simeq 0.3 - 4.0$ at $T/\Tc^{ } = 1.3 - 2.25$.  
In view of table~\ref{table:results}, 
it seems that these values are clearly on the high side, 
as is to be expected from the insufficient resolution of the relativistic
formulation to narrow peaks. 

We end by noting that the cost-effective 
results obtained with our approach
suggest unquenched simulations as a goal for the future, 
whereas within the quenched approximation, 
systematic uncertainties deserve to  
be further scrutinized 
(cf.\ \figs\ref{fig:volume}--\ref{fig:results} and \ref{fig:extrapolation}).

%
\section*{Acknowledgements}

We thank Guy Moore for suggesting that 
appendix~C is worth reporting. 
The computations were performed on the clusters of the
Department of Theoretical Physics, TIFR, the ILGTI, TIFR computing
facilities, as well as 
the Extreme Science and Engineering Discovery
Environment (XSEDE), which is supported by National Science Foundation 
grant number ACI-1548562 (allocation ID: TG-PHY170036).
We would like to thank Ajay Salve and Kapil Ghadiali 
for technical assistance.
S.D.\ acknowledges support of the Department of Atomic Energy, 
Government of India, under Project Identification No.\ RTI 4002.
M.L.\ was supported by the Swiss National Science Foundation
(SNSF), under grant 200020B-188712.

%
\appendix
\renewcommand{\thesection}{\Alph{section}} 
\renewcommand{\thesubsection}{\Alph{section}.\arabic{subsection}}
\renewcommand{\theequation}{\Alph{section}.\arabic{equation}}
%

%
\section{Non-perturbative renormalization of the colour-magnetic operator}
\la{app:rgi}

We review here results for the non-perturbative renormalization 
of the colour-magnetic operator from 
ref.~\cite{renormB}, incorporating minor modifications that
are either necessary for our context, or would not be strictly
necessary, but can be conveniently implemented, given that higher-order 
perturbative results have become available in the meanwhile. 

Ref.~\cite{renormB} considered two possibilities
for the propagation of the heavy quarks in the time direction.
The case with just time-like links appearing, like in 
\eq\nr{GB_def}, corresponds to the Eichten-Hill (EH) action.  

The key result of ref.~\cite{renormB} is how the bare lattice operator
for $[ gB^{ }_i ]^{ }_{\rmi{bare},\rmii{L}}$ from \eq\nr{B_bare} 
is related to the renormalization-group invariant (RGI) 
version of the same operator. This 
is expressed in its \eq(5.7) as 
\be
 Z^\rmii{RGI}_\rmi{spin} 
 = 
 \frac{\Phi^{ }_\rmii{RGI}}{\Phi^{ }_\rmii{SF}(\frac{1}{2 L^{ }_\rmi{max}})}
 \times  
 Z^\rmii{SF}_\rmi{spin} ( 2 L^{ }_\rmi{max} ) 
 \;, \la{eq57}
\ee
where $L^{ }_\rmi{max}$ parametrizes a specific finite-volume 
Schr\"odinger functional (SF) scheme. 
The inverse of the first factor in \eq\nr{eq57}, which is a continuum 
quantity (i.e.\ with no reference to the lattice coupling), 
is given by \eq(5.6) of ref.~\cite{renormB}: 
\be
 \frac{\Phi^{ }_\rmii{SF}(\frac{1}{2L^{ }_\rmi{max}})}{\Phi^{ }_\rmii{RGI}}
 = 0.992 (29)
 \;. \la{eq56}
\ee

It is the second factor in \eq\nr{eq57} which incorporates the 
dependence on the bare lattice coupling. With the step-scaling
approach~\cite{steps}, $Z^\rmii{SF}_\rmi{spin}$ was measured for a number 
of $\beta$-values in ref.~\cite{renormB}. Furthermore,   
an interpolating function was suggested, as 
\be
 Z^\rmii{SF}_\rmi{spin} ( 2 L^{ }_\rmi{max} )
 \stackrel{\rmii{model~1}}{\approx}
 2.58 + 0.14\,(\beta - 6) - 0.27 (\beta-6)^2
 \;, \la{model1}
\ee
where the denomination ``model 1'' is ours.
However, this can only be used in the range $6.0 \le \beta \le 6.5$
where the measurements lie. 
We need to access the range $6.86 \le \beta \le 7.634$, 
and then \eq\nr{model1} cannot be applied. 

To go to larger $\beta$, we can take inspiration from perturbation
theory, even though it is not quantitatively accurate 
in this $\beta$-range. 
One possible representation would
be to consider\footnote{%
 We thank R.~Sommer for suggesting this form. 
 }
\be
 Z^\rmii{SF}_\rmi{spin} ( 2 L^{ }_\rmi{max} )
 \stackrel{\rmii{model~2}}{\simeq} 
 k^{ }_0 - 
 \gamma^{ }_0 
 \ln\Bigl( \frac{\rO}{a} \Bigr)
 \frac{6}{\beta} 
 \;, \la{model2}
\ee
with $k^{ }_0$ treated as a fit parameter and 
the anomalous dimension taking the value
$
 \gamma^{ }_0 = {3}/({8\pi^2})
$.
As another possible fit form, we consider
\be
 Z^\rmii{SF}_\rmi{spin} ( 2 L^{ }_\rmi{max} )
 \stackrel{\rmii{model~3}}{\simeq} 
 k^{ }_1 
 \;, \la{model3}
\ee
which actually provides for a statistically 
better description of the data.

\begin{figure}[t]

\hspace*{-0.1cm}
\centerline{%
   \epsfxsize=7.5cm\epsfbox{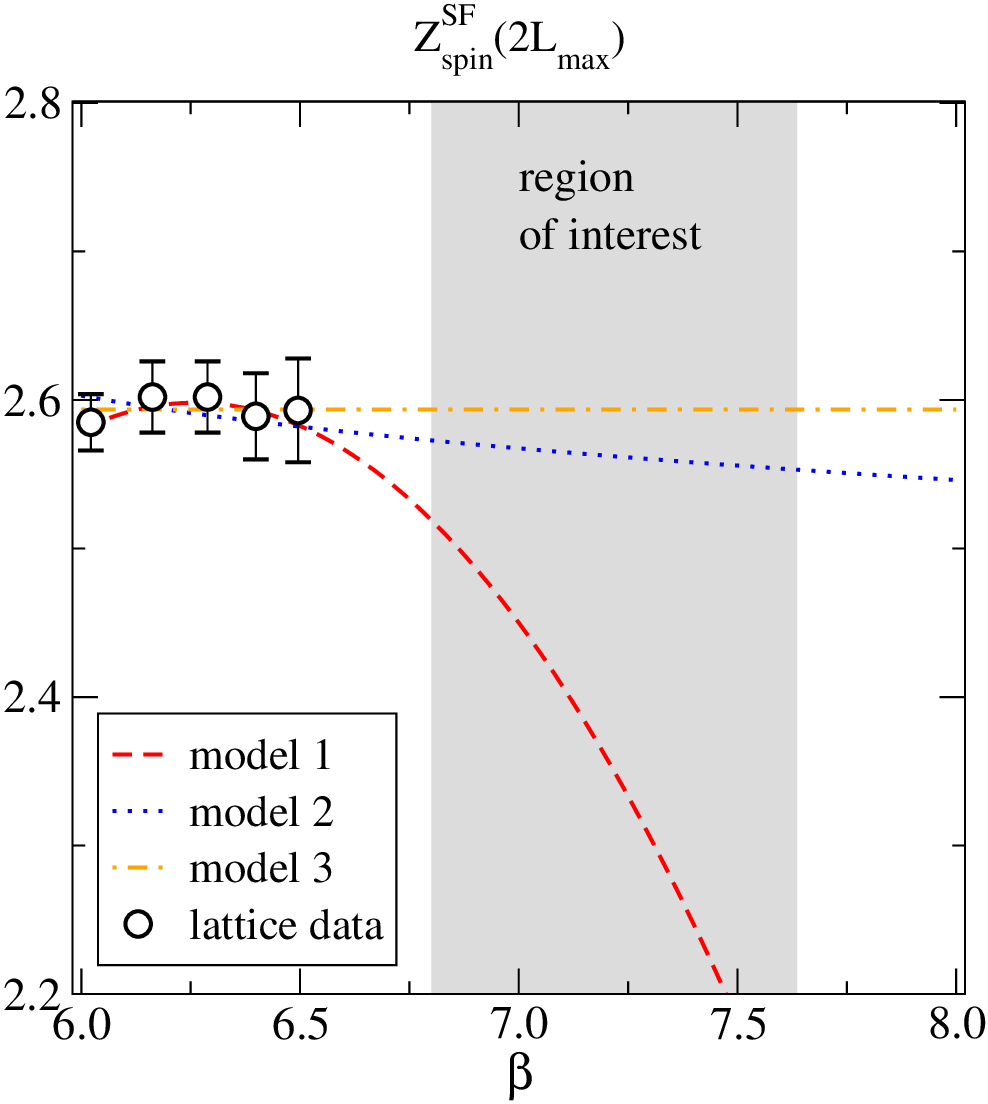}
}

\caption[a]{\small
     Shown are the lattice data from table~3 
     of ref.~\cite{renormB} (case EH), 
     compared with the interpolation from \eq\nr{model1} [model~1] 
     and the extrapolations from \eqs\nr{model2} and \nr{model3}
     [models~2 and~3]. 
     The fit parameters read
     $
      k^{ }_0 \approx 2.66668
     $
     and 
     $
      k^{ }_1 \approx 2.59353
     $.
}

\la{fig:extrapolation}
\end{figure}

The resulting fits, and extrapolation towards large $\beta$, are
illustrated in \fig\ref{fig:extrapolation}. It is clear that in 
the domain $6.86 \le \beta \le 7.634$, \eq\nr{model1} does not 
perform well, as it takes values far away from the 
 interpolation data and has a shape very different from that 
 expected at large $\beta$. 
Eqs.~\nr{model2} and \nr{model3} may also
incorporate substantial systematic uncertainties,  
but the difference of the results obtained with these
two models is much smaller ($\lsim 2\%$) than our statistical
uncertainties (cf.\ \fig\ref{fig:results}). The results 
shown in this paper are based on model~2.

The remaining ingredient is how the RGI operator can be run down to an
$\msbar$ scale. Mirrorring eq.~(5.4) of ref.~\cite{renormB}, 
let us denote this factor by 
\be
 \frac{\Phi^{ }_\rmii{RGI}}{\Phi^{ }_\rmii{$\msbar$}(\bmu)}
 \;=\; 
 [2 b^{ }_0 g^2(\bmu)]^{-\frac{\gamma_0}{2 b_0}}
 \,
 \exp\biggl\{ 
 - 
 \int_0^{g(\bmu)}
 \! {\rm d}g' 
 \biggl[
  \frac{\gamma(g')}{\beta(g')} - \frac{\gamma^{ }_0}{b^{ }_0 g'} 
 \biggr]
 \biggr\} 
 \;.  \la{eq54}
\ee
This factor is a 
function of $\bmu/\Lambdamsbar$, where $\bmu$ is the scale parameter
of the $\msbar$ scheme. Eqs.~(6.7) and (6.8) of ref.~\cite{renormB}
express the resulting conversion as 
\be
 Z^\rmii{$\msbar$}_\rmi{spin}(\bmu) = 
 \frac{\Phi^{ }_\rmii{$\msbar$}(\bmu)}{\Phi^{ }_\rmii{RGI}}
 \times
 Z^\rmii{RGI}_\rmi{spin} 
 \;, \la{eq67}
\ee
citing for 
$
 {\Phi^{ }_\rmii{$\msbar$}(\bmu)} / {\Phi^{ }_\rmii{RGI}}
$
the value 
$ 0.756(18) $ 
for the scale choice 
$
 \frac{\bmu}{\Lambdamsbar} = \frac{\rmii{2~GeV}}{\rmii{238~MeV}} 
$.\footnote{%
 The use of the relations is illustrated a few lines below eq.~(6.8)
 in ref.~\cite{renormB}, 
 suggesting that for $\beta = 6$ and $\bmu = 2$~GeV the overall factor
 from \eqs\nr{eq57}--\nr{model1} and \nr{eq67} should be 
 $
  0.756 \times 0.99 \times 2.58 = 1.93
 $. 
 Unfortunately, there is a bug here, with a comparison
 of \eqs\nr{eq57} and \nr{eq56} indicating that the factor should be
 $
  0.756 \times 2.58 / 0.99 = 1.97
 $.
 We thank H.B.~Meyer and R.~Sommer for confirming our interpretation. 
 } 

For us, the key difference is that we need to run to the $\msbar$
scale $\bmu \approx 19.179 T$, as discussed around 
\eq\nr{thermal_scale}. Recalling that 
$\Tc/\Lambdamsbar = 1.24(10)$~\cite{Tc}, we therefore need to evaluate 
$
   {\Phi^{ }_\rmii{$\msbar$}(\bmu)} / {\Phi^{ }_\rmii{RGI}} 
$ 
with 
$
 {\bmu} / {\Lambdamsbar} \approx
 19.179 \times 1.24 \times \frac{T}{\Tc}
$. This is typically somewhat larger than the scales considered
in ref.~\cite{renormB}, whereby we expect 
$
   {\Phi^{ }_\rmii{$\msbar$}(\bmu)} / {\Phi^{ }_\rmii{RGI}} 
$ 
to be smaller. 

The numerical determination of
$
   {\Phi^{ }_\rmii{$\msbar$}(\bmu)} / {\Phi^{ }_\rmii{RGI}} 
$ 
can in principle be done more 
precisely than in ref.~\cite{renormB}, given that higher order 
corrections to the renormalization group
equations have become available in the meanwhile. 
However, some care is needed in relating the conventions adopted
by the pQCD and lattice communities. 
Denoting 
\be 
 a^{ }_s \;\equiv\; \frac{ \alpha^{ }_s }{ \pi } 
         \;= \;  \frac{g^2}{4\pi^2}
 \;, \quad
 \hat{t} \;\equiv\; \ln\biggl( \frac{\bmu^2}{\Lambda_\tinymsbar^2} \biggr)
 \;, \la{as_def}
\ee
the 5-loop evolution equation for $a^{ }_s$ reads
\be
 \partial^{ }_{\hat{t}}\, a^{ }_s = 
 - (\beta^{ }_0 a_s^2 + ... + \beta^{ }_4 a_s^6)
 \;, \la{rg_as}
\ee
where ($\Nc^{ }=3$, $\Nf^{ }=0$)~\cite{5loop}
\ba
 \beta^{ }_0 & = & \frac{11}{4} 
 \;, \quad
 \beta^{ }_1 \; = \; \frac{51}{8}
 \;, \quad
 \beta^{ }_2 \; = \; \frac{2857}{128}
 \;, \quad 
 \beta^{ }_3 \; = \;  
 \frac{149753}{1536} + \frac{891\zeta^{ }_3}{64}
 \;, \\[2mm]
 \beta^{ }_4 & = & 
 \frac{8157455}{16384} + 
 \frac{621885\zeta^{ }_3}{2048} - 
 \frac{88209\zeta^{ }_4}{2048} - 
 \frac{144045\zeta^{ }_5}{512}
 \;. 
\ea
The initial condition can be fixed 
at large $\hat{t} \equiv \hat{t}^{ }_\rmi{max}$, 
e.g.\ $\hat{t}^{ }_\rmi{max} = 200$, as 
\be
 a^{ }_s(\hat{t}^{ }_\rmi{max}) 
 = 
 \frac{1}{\beta^{ }_0 \hat{t}^{ }_\rmi{max}}
 - 
 \frac{\beta^{ }_1 \ln(\hat{t}^{ }_\rmi{max})}
 {\beta_0^3 \hat{t}_\rmi{max}^2}
 + 
 \frac{\beta^{ }_0 \beta^{ }_2 + \beta_1^2\,
 [\ln^2(\hat{t}^{ }_\rmi{max}) - \ln(\hat{t}^{ }_\rmi{max}) - 1]}
 {\beta_0^5 \hat{t}_\rmi{max}^3}
 + 
 \rmO\biggl( \frac{1}{\hat{t}_\rmi{max}^4} \biggr)
 \;, \la{as_initial}
\ee
which solves \eq\nr{rg_as} up to 3-loop order, 
and guarantees that $\Lambdamsbar$ has its standard meaning.

To relate these to \eq\nr{eq54}, we note that 
in \eq\nr{eq54} the evolution 
equation for the running coupling has been assumed to have the form
\be
 \bmu \frac{{\rm d} g}{{\rm d}\bmu} = \beta(g) 
 = -g^3 \, \bigl( b^{ }_0 + b^{ }_1 g^2 + ... \bigr)
 \;. \la{rg_as_alt}
\ee
A comparison with \eq\nr{rg_as} yields
\be
 b^{ }_0 = \frac{\beta^{ }_0}{4\pi^2}
 \;, \quad
 b^{ }_1 = \frac{\beta^{ }_1}{(4\pi^2)^2}
 \;, \quad ...
 \;. 
\ee

Now, making use of \eq\nr{rg_as_alt}, it is straightforward to verify that
the ratio in \eq\nr{eq54} satisfies
\be
  \bmu \frac{{\rm d} }{{\rm d}\bmu}
 \biggl[ 
  \frac{\Phi^{ }_\rmii{RGI}}{\Phi^{ }_\rmii{$\msbar$}(\bmu)}
 \biggr]
 \; = \; 
 -\gamma(g) \,  
 \biggl[ 
  \frac{\Phi^{ }_\rmii{RGI}}{\Phi^{ }_\rmii{$\msbar$}(\bmu)}
 \biggr]
 \;, \la{rg_zcm}  
\ee
where $\gamma(g)$ has the expansion 
\be
 \gamma(g) = -g^2 \bigl( \gamma^{ }_0 + \gamma^{ }_1 g^2 + ... \bigr)
 \;. \la{gamma_g}
\ee
In the notation of ref.~\cite{hqet3}, the corresponding quantity
is $\gamma^{ }_\rmi{cm}$, given in its \eq(13), 
\be
 \gamma^{ }_\rmi{cm} 
 = \tilde\gamma^{ }_0\, a^{ }_s + \tilde \gamma^{ }_1\, a_s^2
    + \tilde \gamma^{ }_2\, a_s^3 + ...  
 \;,  
\ee
where  ($\Nc^{ }=3$, $\Nf^{ }=0$)~\cite{hqet3}
\ba
 \tilde\gamma^{ }_0 & = & \frac{3}{2} 
 \;, \quad
 \tilde\gamma^{ }_1 \; = \; \frac{17}{4}
 \;, \quad
 \tilde\gamma^{ }_2 \; = \; \frac{899}{64} + \frac{15\pi^2}{16}
 + \frac{27\zeta^{ }_3}{8}
 \;. \la{tilde_gamma}
\ea
Recalling the definition of $a_s^{ }$ from \eq\nr{as_def}, 
the relation of the couplings in \eqs\nr{gamma_g} and \nr{tilde_gamma} is 
\be
 \gamma^{ }_0 = \frac{\tilde\gamma^{ }_0}{4\pi^2}
 \;, \quad
 \gamma^{ }_1 = \frac{\tilde\gamma^{ }_1}{(4\pi^2)^2}
 \;, \quad ... 
 \;. 
\ee

\begin{figure}[t]

\hspace*{0.4cm}
\begin{minipage}[c]{7.5cm}
\centerline{%
   \epsfxsize=7.5cm\epsfbox{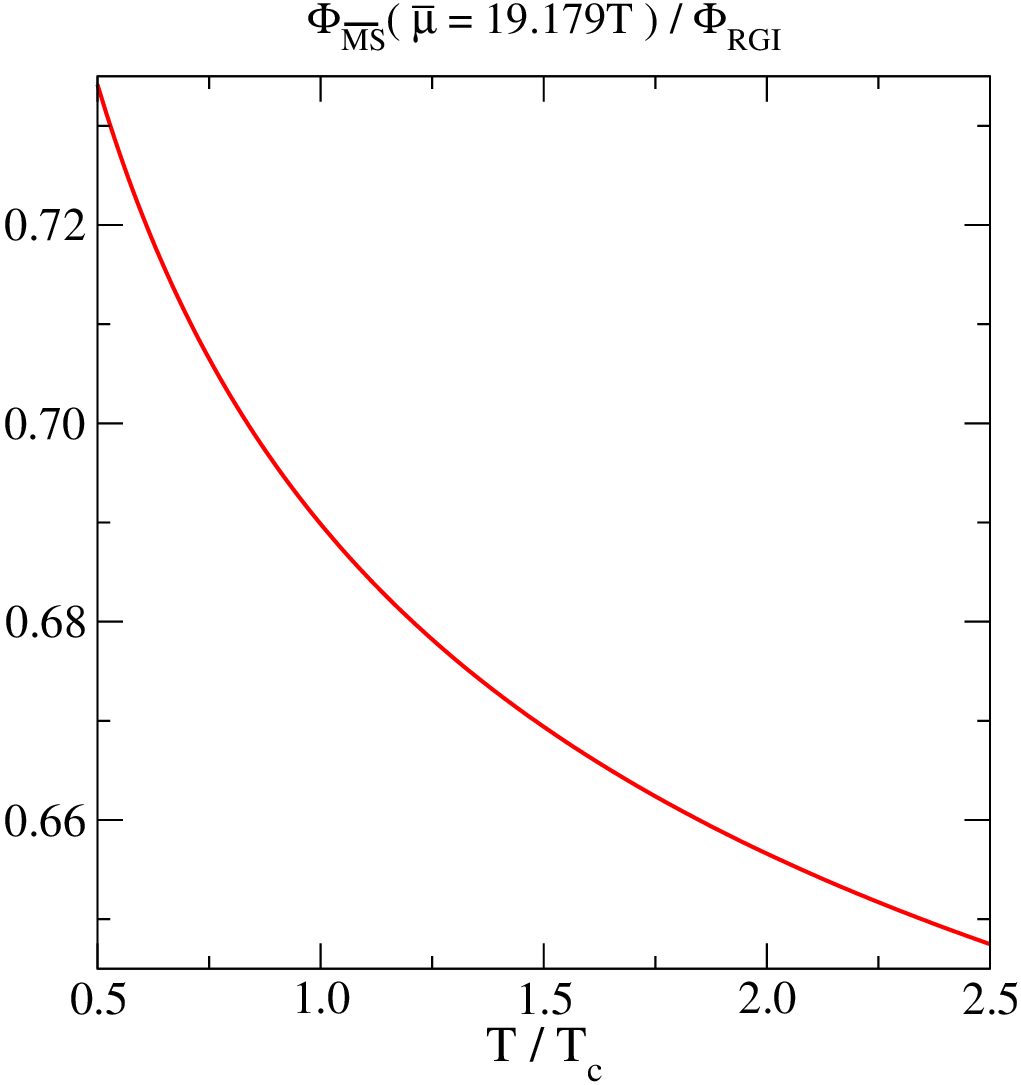}
}
\end{minipage}
\begin{minipage}[c]{7.0cm}
\small{
\begin{center}
\vspace*{-4mm}
\begin{tabular}{cc} 
 $ T / \Tc^{ } $ & 
 ${\Phi^{ }_\rmii{$\msbar$}}(\bmu = 19.179T)/
  {\Phi^{ }_\rmii{RGI}}$
 \\[2mm]
 \hline
 \\[-4mm] 
  0.6 & 0.7210
  \\
  0.8 & 0.7026
  \\
  1.0 & 0.6898
  \\
  1.2 & 0.6803
  \\
  1.4 & 0.6727
  \\
  1.5 & 0.6694
  \\
  1.6 & 0.6664
  \\
  1.8 & 0.6611
  \\
  2.0 & 0.6566
  \\
  2.2 & 0.6526
  \\
  2.4 & 0.6491
  \\
 \hline 
\end{tabular} 
\end{center}
}
\end{minipage}

\caption[a]{\small
     Illustration of the conversion factor
         ${\Phi^{ }_\rmii{$\msbar$}}/
          {\Phi^{ }_\rmii{RGI}}$, 
     needed for pulling back an RGI-renormalized colour-magnetic field
     to the $\msbar$ scale $ \bmu \approx 19.179 T $ that is need
     in the thermal context~(cf.\ \eq\nr{thermal_scale}). 
}

\la{fig:zcm}
\end{figure}

The initial condition for 
\eq\nr{rg_zcm} can be obtained from a perturbative solution 
of \eq\nr{eq54}. 
Denoting 
$
 g^2(\bmu^{ }_\rmi{max}) \equiv 4\pi^2 a^{ }_s(\hat{t}^{ }_\rmi{max})
$, 
where $a^{ }_s(\hat{t}^{ }_\rmi{max})$  is from \eq\nr{as_initial}, 
this reads
\be
 \frac{\Phi^{ }_\rmii{RGI}}{\Phi^{ }_\rmii{$\msbar$}(\bmu^{ }_\rmi{max})}
 \;\approx \; 
 [2 b^{ }_0 g^2(\bmu^{ }_\rmi{max})]^{-\frac{\gamma_0}{2 b_0}}
 \,
 \exp\biggl\{ 
 - \frac{\gamma^{ }_0}{2 b^{ }_0}
 \biggl(
  \frac{\gamma^{ }_1}{\gamma^{ }_0} - \frac{b^{ }_1}{b^{ }_0} 
 \biggr)
 \, 
 g^2(\bmu^{ }_\rmi{max})
 \, 
 \biggr\} 
 \;.  \la{eq54_max}
\ee

Solving the coupled set of differential equations from 
\eqs\nr{rg_as} and \nr{rg_zcm}, 
with the initial conditions from \eqs\nr{as_initial} and \nr{eq54_max}, 
and plotting the inverse of the result, as needed in \eq\nr{eq67}, 
we obtain the curve shown in \fig\ref{fig:zcm}. This is the result
employed in our analysis. 

%
\section{Lattice perturbation theory and tree-level improvement}
\la{app:lpt}

We have computed the correlator obtained with the discretization 
of \eqs\nr{B_bare}--\nr{plaquette} 
to leading order in lattice perturbation theory. 
The result differs slightly from the corresponding result for the 
colour-electric correlator~\cite{lat25}, even if the 
continuum limit is the same. Denoting by $\Ntau$ the number of 
lattice points in the time direction, the new result reads  
\ba
 && \hspace*{-8mm}
 \bigl[ G_{\B}^{ }(\tau) \bigr]^\rmii{LO}_{\rmi{bare},\rmii{L}}
 =
 \frac{g_\rmi{0}^2 \CF}{3a^4} 
 \! \mathop{\int}_{-\pi}^{\pi} \! 
 \frac{{\rm d}^3\vec{q}}{(2\pi)^3} \, 
 \frac{\mathrm{e}^{\bar{q}\Ntau(1-\tau T)}
 +\mathrm{e}^{\bar{q}\Ntau\tau T}}
 {\mathrm{e}^{\bar{q}\Ntau}-1}
 \;
 \frac{\tilde{q}^2 - \frac{ (\tilde{q}^2)^2 + \tilde{q}^4 }{8} 
 + \frac{  \tilde{q}^2 \tilde{q}^4 - \tilde{q}^6 }{32}  } 
 {\mathrm{sinh}\;{\bar{q}}}
 \;,  \hspace*{6mm}  \la{GB_lat} \\ 
 && \hspace*{-8mm} 
 \bar{q} \; \equiv \; 2\;\mathrm{arsinh}
 \Bigl(\frac{\sqrt{\tilde{q}^2}}{2}\Bigr)
 \;, \quad 
 \tilde{q}^n\;\equiv\; \sum_{i=1}^3 2^n \sin^n{\left(\frac{q_i}{2}\right)}
 \;. \hspace*{3mm}
\ea
At this order the coupling $g_\rmi{0}^2$ 
denotes the bare lattice coupling, 
$g_\rmi{0}^2 \equiv 6 / \beta$. For purposes of normalization, 
it is convenient to eliminate the dependence on the coupling, 
whence we define
\be
 G^\rmii{ }_{\rmi{norm},\rmii{L}} (\tau) 
 \; \equiv \; 
 \frac{ 
 \bigl[ G_{\B}^{ }(\tau) \bigr]^\rmii{LO}_{\rmi{bare},\rmii{L}}
  }{
  g_\rmi{0}^2 \CF
  }
 \;. \la{G_norm_L} 
\ee
The corresponding continuum correlator is given in \eq\nr{G_norm_DR}. 

\begin{figure}[t]

\hspace*{-0.1cm}
\centerline{%
   \epsfxsize=7.5cm\epsfbox{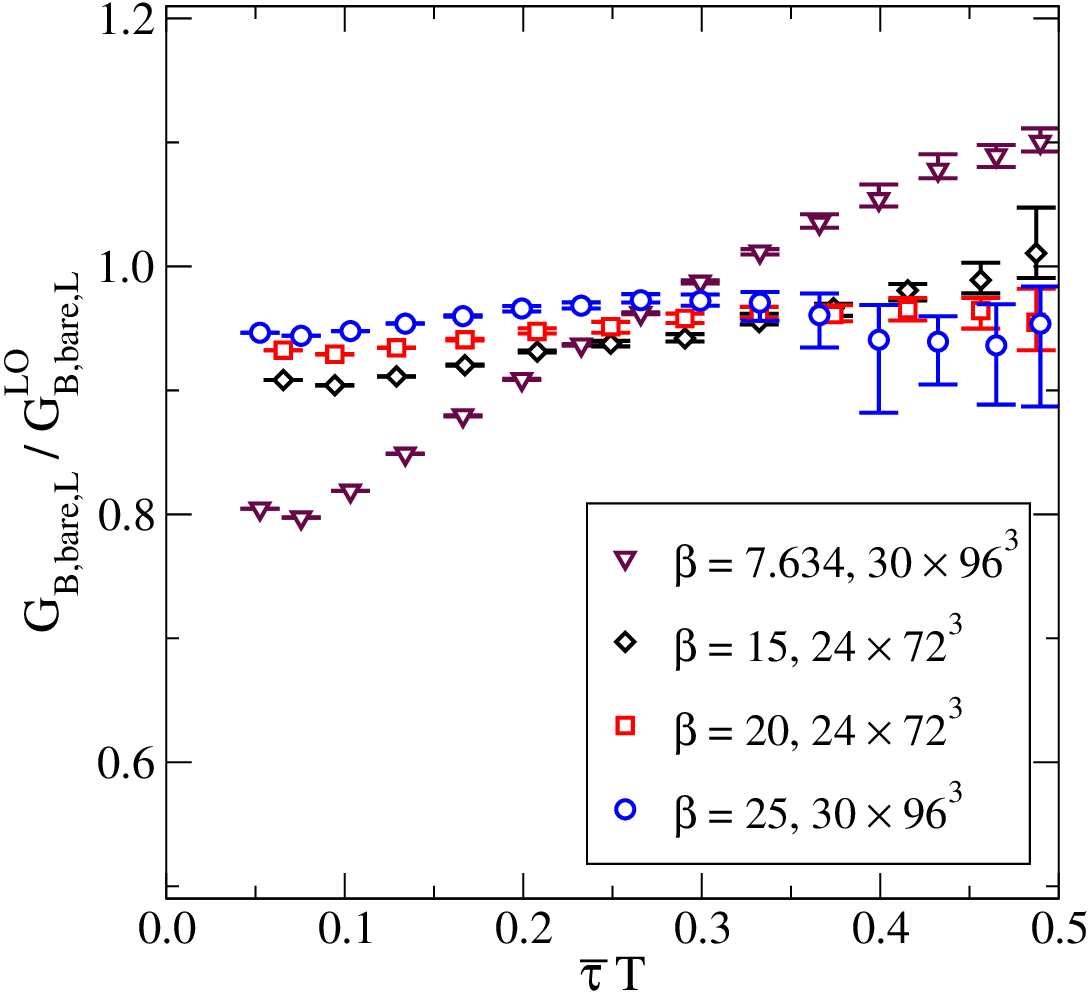}
}

\caption[a]{\small
     Approach of the lattice measurement towards the 
     tree-level prediction in \eq\nr{GB_lat}. Shown is one of our
     production sets 
     (without renormalization from \eq\nr{master}, 
     which would be $\{...\}^2_{ }\approx 2.871$) 
     and, for comparison, three short runs in which  
     the bare coupling 
     $g_\rmi{0}^2 = 6/\beta$ was made 
     extremely small. 
}

\la{fig:crosscheck}
\end{figure}

Eqs.~\nr{G_norm_DR} and \nr{G_norm_L} permit to implement
``tree-level improvement''~\cite{rs,hbm}, which can significantly 
reduce discretization effects at small distances.
Using \eqs\nr{G_norm_DR} and \nr{G_norm_L}
to determine $\overline{\tau}$ from
\be
 G^\rmii{ }_{\rmi{norm},\rmii{DR}} ( \overline{\tau} )
 \; \equiv \; 
 G^\rmii{ }_{\rmi{norm},\rmii{L}} (\tau) 
 \;, \la{improvement}
\ee
the tree-level improved correlator is obtained from  
\be
 \bigl[ G_{\B}^{ }( \overline{\tau} ) \bigr]^\rmi{imp}_{\rmi{bare},\rmii{L}}
 \; \equiv \; 
 \bigl[ G_{\B}^{ }(\tau) \bigr]^{ }_{\rmi{bare},\rmii{L}}
 \;,
\ee
i.e.\ the measured values at distance $\tau$ 
are assigned to a corrected value $\overline{\tau}$. 

Another use of the perturbative expression in \eq\nr{GB_lat}
is that it permits to crosscheck 
the overall normalization of the lattice measurement, 
by going to very large $\beta$.
Results are shown 
in \fig\ref{fig:crosscheck}, and at short separations
 display a gradual movement towards
 unity as $\beta$ increases, thereby confirming that everything is 
 in order.

%
\section{Colour-magnetic spectral function at next-to-leading order}
\la{app:nlo}

%
\begin{figure}[t]
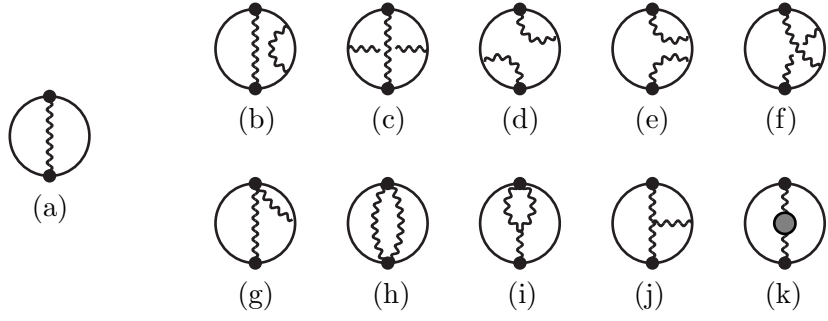


\hspace*{1.5cm}%
\begin{minipage}[c]{3cm}
\begin{eqnarray*}
&& 
 \hspace*{-1cm}
 \EleA 
\\[1mm] 
&& 
 \hspace*{-0.6cm}
 \mbox{(a)} 
\end{eqnarray*}
\end{minipage}%
\begin{minipage}[c]{10cm}
\begin{eqnarray*}
&& 
 \hspace*{-1cm}
 \EleB \quad\; 
 \EleBB \quad\; 
 \EleC \quad\; 
 \EleD \quad\; 
 \EleE \quad\; 
\\[1mm] 
&& 
 \hspace*{-0.6cm}
 \mbox{(b)} \hspace*{1.26cm}
 \mbox{(c)} \hspace*{1.26cm}
 \mbox{(d)} \hspace*{1.26cm}
 \mbox{(e)} \hspace*{1.26cm}
 \mbox{(f)} 
\\[5mm] 
&& 
 \hspace*{-1cm}
 \EleF \quad\; 
 \EleG \quad\; 
 \EleH \quad\; 
 \EleI \quad\; 
 \EleJ \quad 
\\[1mm] 
&& 
 \hspace*{-0.6cm}
 \mbox{(g)} \hspace*{1.3cm}
 \mbox{(h)} \hspace*{1.28cm}
 \mbox{(i)} \hspace*{1.28cm}
 \mbox{(j)} \hspace*{1.32cm}
 \mbox{(k)} 
\end{eqnarray*}
\end{minipage}

\caption[a]{\small 
  The Feynman diagrams contributing to the continuum version
  of the colour-magnetic correlator defined in \eq\nr{GB_def}. 
  The circle illustrates a Polyakov loop, the wavy lines are 
  gluons, the small dots stand for insertions of $B^{ }_i$, 
  and the grey blob is the gluon self-energy. } 
\la{fig:GB}
\end{figure}
%

At next-to-leading order, the continuum diagrams contributing 
to $ G^{ }_{\!\B}(\tau) $ are those in \fig\ref{fig:GB}.\footnote{%
 To simplify the notation
 we omit the specifier $[...]^{ }_{\rmi{bare},\rmii{DR}}$
 in this section. 
 } 
The leading-order contribution reads
\be
 \delta^{ }_\rmi{(a)} G^{ }_{\!\B}(\tau)  
 = \frac{\gB^2 \CF^{ }(D-3)(D-2)}{3} \, \J^{ }_0(\tau) 
 \;,  \la{GB_LO}
\ee
where 
$D \equiv 4 - 2\epsilon$ is the space-time dimension, 
$\J^{ }_0$ is defined in \eq\nr{def_J0}, 
and the bare coupling reads
\be
 \gB^2 = g^2 + \frac{g^4 \mu^{-2\epsilon}}{(4\pi)^2} 
 \frac{2\Nf - 11\Nc}{3\epsilon} + \rmO(g^6)
 \;. \la{gB}
\ee

At next-to-leading order, 
following a numbering for ``master'' 
structures adopted from refs.~\cite{rhoE,1overM},
the non-vanishing contributions of the individual graphs read
\ba
 & & \hspace*{-2.5cm} 
 \delta^{ }_\rmi{(a)}
 G^{ }_{\!\B}(\tau)
 \times 
 \frac{\chi^{ }_\rmiii{LO}}{\chi^{ }_\rmiii{NLO}}
 + 
 \delta^{ }_\rmi{(b-f)}
 G^{ }_{\!\B}(\tau)
 \; = \;
 \frac{g^4 \Nc^{ } \CF^{ }(D-3)(D-2)}{3} \, \I^{ }_4(\tau)
 \;, 
 \\[2mm]
 \delta^{ }_\rmi{(h)}
 G^{ }_{\!\B}(\tau) & = & 
 \frac{g^4 \Nc^{ } \CF^{ }(D-3)(D-2)(D-1)}{6} \, \I^{ }_1(\tau)
 \;,
 \\[2mm] 
 \delta^{ }_\rmi{(i)}
 G^{ }_{\!\B}(\tau) & = & - 
 g^4 \Nc^{ } \CF^{ }(D-3)(D-2) \, \I^{ }_2(\tau)
 \;, 
 \\[2mm]
 \delta^{ }_\rmi{(j)}
 G^{ }_{\!\B}(\tau) & = & 
 \frac{g^4 \Nc^{ } \CF^{ }(D-3)(D-2)}{6} \, \I^{ }_6(\tau)
 \;, 
 \\[2mm]
 \delta^{ }_\rmi{(k)}
 G^{ }_{\!\B}(\tau) & = & 
 \frac{g^4 \CF^{ }(D-3)}{3} \, \biggl\{ 
  (D-2)\Nc^{ }
  \biggl[
  - \frac{(D-2)(D-1)}{2}\, \I^{ }_0(\tau)
  + 2\, \I^{ }_2(\tau) 
  + 2\, \I^{ }_7(\tau)
  \biggr]
 \nn 
 & & 
 + \, \Nf^{ }
  \biggl[
     (D-2)(D-1)\, \I^{ }_{\{0\}}(\tau) 
   - (D-2)\, \I^{ }_{\{2\}}(\tau)
   - 4\, \I^{ }_{\{7\}}(\tau)
  \biggr]
  \biggr\}
 \;. 
\ea
Here $\chi$ refers to the denominator of \eq\nr{GB_def}. 

As Matsubara sum-integrals, or in terms of a configuration-space 
representation in the case of \eq\nr{def_I6}, the master structures 
are defined as 
\ba
 \J^{ }_0(\tau) & \equiv &
 \Tint{K}
 \frac{k^2 e^{i k_n\tau}}{K^2} 
 \;,
 \la{def_J0} 
 \\[2mm] 
 \I^{ }_0(\tau) & \equiv &
 \Tint{K,Q}
 \frac{e^{i k_n\tau}}{K^2(K-Q)^2} 
 \;, 
 \la{def_I0}
 \\[2mm] 
 \I^{ }_1(\tau) & \equiv &
 \Tint{K,Q}
 \frac{e^{i k_n\tau}}{Q^2(K-Q)^2}
 \;, 
 \la{def_I1} 
 \\[2mm] 
 \I^{ }_2(\tau) & \equiv &
 \Tint{K,Q} 
 \frac{k^2 e^{i k_n\tau}}{K^2 Q^2 (K-Q)^2}
 \;, 
 \la{def_I2}
 \\ 
 \I^{ }_4(\tau) & \equiv & 
 \Tint{K} \frac{k^2 e^{i k_n\tau}}{K^2}
 \Tint{Q'} \frac{e^{i q_n\tau} - 1}{q_n^2 Q^2} 
 \;, 
 \la{def_I4}
 \\[2mm] 
 \I^{ }_6(\tau) & \equiv & 
 \biggl[
  \int_{\tau}^{\beta} \! {\rm d}\tau'
    - 
  \int_{0}^{\tau} \! {\rm d}\tau' 
 \biggr] 
 \int_X
   G(X - \tau'\vec{e}^{ }_0 ) 
  \nn 
  & \times &  
   \Bigl[ 
          \partial^{ }_i G(X - \tau \vec{e}^{ }_0) 
          \,\partial^{ }_0 \partial^{ }_i  G(X) 
                - 
          \partial^{ }_0 \partial^{ }_i G(X-\tau \vec{e}^{ }_0) 
          \,\partial^{ }_i G(X) 
   \Bigr]
 \;, \hspace*{4mm} 
 \la{def_I6}
 \\[2mm] 
 \I^{ }_7(\tau) & \equiv & - \lim_{\lambda\to 0}
 \frac{{\rm d}}{{\rm d}\lambda^2}
 \Tint{K,Q} 
 \frac{[k^2 q^2 - (\vec{k}\cdot\vec{q}\,)^2 ] e^{i k_n\tau}}
      {(K^2+\lambda^2)Q^2(K-Q)^2} 
 \;, \la{def_I7}
\ea
where for \eq\nr{def_I6} we have defined
$
 G(X)
 \, \equiv \, 
 \Tinti{K}\, {e^{i K\cdot X}} / {K^2}
$.
Moreover, the notation $\I^{ }_{\{ i \}}$ implies that the thermal 
distribution is fermionic (cf.\ the line after \eq\nr{tildeI7}).

To obtain the spectral function, we go over to frequency space, 
carry out an analytic continuation, and take the cut, 
\be
 \tilde\I^{ }_i(\omega) 
 \; \equiv \; 
 \im \biggl[ 
  \int_0^{1/T} \! {\rm d}\tau \, e^{i k^{ }_n \tau} \, \I^{ }_i(\tau) 
     \biggr]^{ }_{ k^{ }_n \to -i [\omega + i 0^+_{ }] }
 \;. 
\ee
Employing techniques explained in ref.~\cite{rhoE}, 
this leads to the integral representations 
\ba
 \tilde{\mathcal{J}}^{ }_0(\omega) & = & 
 \frac{\omega^3\mu^{-2\epsilon}}{4\pi}
 \biggl[
 1 + \epsilon \,  
 \biggl( \ln\frac{\bmu^2}{4\omega^2} + 2 
 \biggr) 
 \biggr]
 + \rmO(\epsilon^2) 
 \;,  
 \\[2mm]
 \tilde{\mathcal{I}}^{ }_0(\omega) & = & 
 \frac{1}{16\pi^3} \int_0^\infty \! {\rm d} q \, n(q) 
 \bigl[
   2 q \omega 
 \bigr] 
 + \rmO(\epsilon) 
 \;, 
 \la{tildeI0} 
 \\[2mm]
 \tilde{\mathcal{I}}^{ }_1(\omega) & = & 
 \frac{\omega^3\mu^{-4\epsilon}}{16\pi^3}
 \biggl( \frac{1}{6}
 \biggr)  
 + \frac{1}{16\pi^3} \int_0^\infty \! {\rm d} q \, n(q) 
 \bigl[
   4 q \omega 
 \bigr]  
 + \rmO(\epsilon) 
 \;, 
 \la{tildeI1} 
 \\[2mm]
 \tilde{\mathcal{I}}^{ }_2(\omega) & = & 
 \frac{\omega^3\mu^{-4\epsilon}}{16\pi^3}
 \biggl( \frac{1}{4\epsilon} + \fr12 \ln\frac{\bmu^2}{4\omega^2} + \fr{5}3 
 \biggr)  
 \nn & + &  
 \frac{1}{16\pi^3} \int_0^\infty \! {\rm d} q \, n(q) 
 \biggl[
   4 q \omega + \omega^2 \ln \left| \frac{q+w}{q-w} \right| 
 \biggr]  
 + \rmO(\epsilon) 
 \;,  
 \la{tildeI2} 
 \\[2mm]
 \tilde{\mathcal{I}}^{ }_4(\omega) & = & 
 \frac{\omega^3\mu^{-4\epsilon}}{16 \pi^3}
 \biggl( \frac{1}{2\epsilon} + \ln\frac{\bmu^2}{4\omega^2} + \fr{23}{6} 
 \biggr)  
 \nn & - &  
 \frac{1}{16\pi^3} \int_0^\infty \! {\rm d} q \, n(q) 
 \biggl[
   4 q \omega + 
  \mathbbm{P}\biggl( \frac{2q \omega^3 }{\omega^2 - q^2} \biggr)
 \biggr]  
 + \rmO(\epsilon) 
 \;, 
 \la{tildeI4} 
 \\[2mm]
 \tilde\I^{ }_6(\omega) & = & 
 \frac{\omega^3 \mu^{-4\epsilon}}{16\pi^3}
 \biggl(
  \frac{1}{2\epsilon} + \ln\frac{\bmu^2}{4\omega^2} + \frac{16}{3}
  - \frac{4\pi^2}{3}
 \biggr)
 \nn & + &  
 \frac{1}{8\pi^3}
 \int_0^\infty \! {\rm d}q \, n(q) \, \mathbbm{P} \biggl\{ 
    4 q \omega \biggl( 1 + \frac{\omega^2}{\omega^2 - q^2} \biggr) 
  + 5 \omega^2 \ln \biggl| \frac{q+\omega}{q-\omega} \biggr|
  \nn 
 & + & 
   \frac{2\omega^4}{q}
   \biggl[
    \frac{1}{q+\omega} \ln \frac{q+\omega}{\omega} -  
    \frac{1}{q-\omega} \ln \frac{|q-\omega|}{\omega}
   \biggr]
 \biggr\}  
 + \rmO(\epsilon) 
 \;,
 \la{tildeI6}
 \\[2mm]
 \tilde\I^{ }_7(\omega) 
 & = & 
 \frac{\omega^3 \mu^{-4\epsilon}}{16\pi^3}
 \biggl(
  -\frac{1}{24\epsilon} 
  -\frac{1}{12} \ln\frac{\bmu^2}{4\omega^2}
 - \frac{19}{72}
 \biggr)
 \nn & + & 
 \frac{1}{16\pi^3}
 \int_0^{\infty} \! {\rm d}q \, n(q) \, 
 \biggl\{ 
   q^2 \ln\biggl| \frac{q + \omega}{q - \omega} \biggr|
  + q \omega \ln \biggl| \frac{q^2 - \omega^2}{\omega^2} \biggr|
 \biggr\}  
 + \rmO(\epsilon) 
 \;,
 \la{tildeI7}
\ea
where $\mathbbm{P}$ denotes a principal value, and 
$n \equiv \nB^{ }$ for bosons and $ n \equiv -\nF^{ } $ for fermions. 

Inserting the integral representations into the individual terms, 
summing them together, 
and going over to the $\msbar$ scheme according to \eq\nr{msbar}, 
the final result becomes
\ba
 \bigl[ \rho^{ }_\B (\omega) \bigr]^{ }_\rmi{renorm,$\bmu$} & = & 
 \frac{g^2 \CF^{ }\, \omega^3}{6\pi} 
 \nn & \times & 
 \biggl\{ 1 + 
 \frac{g^2}{(4\pi)^2}
 \biggl[
      \Nc \biggl( 
        \frac{5}{3} \ln\frac{\bmu^2}{4\omega^2} 
      + \frac{134}{9} - \frac{8\pi^2}{3}
          \biggr) 
  -  \Nf  \biggl(
         \fr23 \ln\frac{\bmu^2}{4\omega^2} + \frac{26}{9}
         \biggr)
 \biggr] 
 \biggr\} 
 \nn[2mm] & + &  
 \frac{g^4 \CF^{ }}{12 \pi^3}
 \biggl\{ 
 \; 
 \Nc \int_0^\infty \! {\rm d} q \, \nB^{ }(q)\, \mathbbm{P}
 \biggl[
   (q^2 +2 \omega^2) \ln \left| \frac{q+w}{q-w} \right| 
  \nn  & & \hspace*{3cm} + \; 
   q \omega \biggl( \ln\frac{|q^2 - \omega^2|}{\omega^2}
   + \frac{\omega^2}{\omega^2 - q^2}
   - 2 \biggr)
  \nn & & \hspace*{3cm} + \;
     \frac{\omega^4}{q} 
       \biggl( \frac{1}{q+\omega} \ln \frac{q + \omega}{\omega} 
     +   \frac{1}{q-\omega} \ln \frac{\omega}{|q - \omega|} \biggr) \biggr]
 \nn[2mm] & & \hspace*{0.9cm} + \; 
 \Nf^{ } \int_0^\infty \! {\rm d} q \, \nF^{ }(q) 
 \biggl[
   \Bigl(q^2 +\frac{\omega^2}{2}\Bigr) \ln \left| \frac{q+w}{q-w} \right| 
  \nn  & & \hspace*{3cm} + \; 
   q \omega \biggl( \ln\frac{|q^2 - \omega^2|}{\omega^2} - 1 \biggr)\biggr] 
 \biggr\} 
  \; + \; \rmO\bigl(g^6\bigr) \;. \hspace*{0.7cm}
 \la{rhoB_nlo}
\ea
The ``physical'' spectral function, 
appearing in~\eq\nr{rhoB_def}, 
involves a multiplication of this expression
through $c_\B^2$, cf.\ \eq\nr{def_GB_physical}. 
The IR limit agrees with an expression given in ref.~\cite{1overM}, 
\ba
 \rho^{ }_\B (\omega) 
  & \stackrel{\omega \ll \pi T}{\approx}  & 
 \frac{g^4 \CF^{ } T^2 \omega }{36\pi}
 \biggl\{ 
  \Nc^{ }\, 
            \biggl[
  \ln\biggl(\frac{T}{\omega}\biggr) 
  + 1 - \gammaE + \frac{\zeta'(2)}{\zeta(2)} 
            \biggr]
 \nn 
 &  &
 \hspace*{1.5cm} + \, 
  \frac{\Nf^{ }}{2}
            \biggl[ 
  \ln\biggl(\frac{2T}{\omega}\biggr)
  + \frac{3}{2} - \gammaE + \frac{\zeta'(2)}{\zeta(2)} 
            \biggr]
 \biggr\} 
 \; + \; \rmO(g^6) 
 \;. \hspace*{0.5cm} \la{rhoB_IR}
\ea

%
\section{Details of spectral fitting procedure}
\la{app:fits}

%
\begin{table}[t]

\small{
\begin{center}
\begin{tabular}{cccccc} 
 fit form & 
 $N^{ }_\tau = 20$ &  
 $N^{ }_\tau = 24$ &  
 $N^{ }_\tau = 28$ &  
 $N^{ }_\tau = 30$ &  
 $N^{ }_\tau\to \infty $
 \\[2mm]
 \hline
 \\[-4mm] 
 \eq\nr{model_B} &
      1.30 -- 1.94 & 1.20 -- 1.75
    & 1.09 -- 1.82 & 1.56 -- 2.00 & 1.56 -- 1.82
  \\[0.5mm]  
 \eq\nr{formc} &
      1.63 -- 2.23 & 1.50 -- 2.26
    & 1.44 -- 2.28 & 2.01 -- 2.59 & 1.79 -- 2.30
 \\[0.5mm]  
 \eq\nr{fourierb} & 
      1.30 -- 1.90 & 1.20 -- 1.66
    & 1.05 -- 1.67 & 1.46 -- 2.01 & 1.36 -- 1.75
 \\[0.5mm]  
 \eq\nr{fourierc} &
      1.56 -- 2.19 & 1.43 -- 2.13
    & 1.30 -- 2.18 & 1.88 -- 2.51 & 1.71 -- 2.14
 \\  
 \hline 
\end{tabular} 
\end{center}
}


\caption[a]{\small
 Fit results for $\kappa^{ }_{\B}/T^3$ 
 at $T = 1.2 \Tc^{ }$, 
 as described in appendix~D.  
 }
\label{table:1.2Tc}
\end{table}
%

%
\begin{table}[t]

\small{
\begin{center}
\begin{tabular}{ccccc} 
 fit form & 
 $N^{ }_\tau = 20$ &  
 $N^{ }_\tau = 24$ &  
 $N^{ }_\tau = 28$ &  
 $N^{ }_\tau\to \infty $
 \\[2mm]
 \hline
 \\[-4mm] 
 \eq\nr{model_B} &
      1.31 -- 1.76 & 1.22 -- 1.64
    & 1.15 -- 1.67 & 1.18 -- 1.67
  \\[0.5mm]  
 \eq\nr{formc} &
      1.70 -- 2.09 & 1.41 -- 2.05
    & 1.38 -- 2.10 & 1.32 -- 1.87
 \\[0.5mm]  
 \eq\nr{fourierb} & 
      1.23 -- 1.68 & 1.05 -- 1.49
    & 1.16 -- 1.64 & 1.13 -- 1.38
 \\[0.5mm]  
 \eq\nr{fourierc} &
      1.60 -- 1.99 & 1.36 -- 1.92
    & 1.25 -- 1.94 & 1.31 -- 1.81
 \\  
 \hline 
\end{tabular} 
\end{center}
}


\caption[a]{\small
 Fit results for $\kappa^{ }_{\B}/T^3$ 
 at $T = 1.5 \Tc^{ }$, 
 as described in appendix~D.  
 }
\label{table:1.5Tc}
\end{table}
%

\begin{figure}[t]

\hspace*{-0.1cm}
\centerline{%
   \epsfxsize=5.0cm\epsfbox{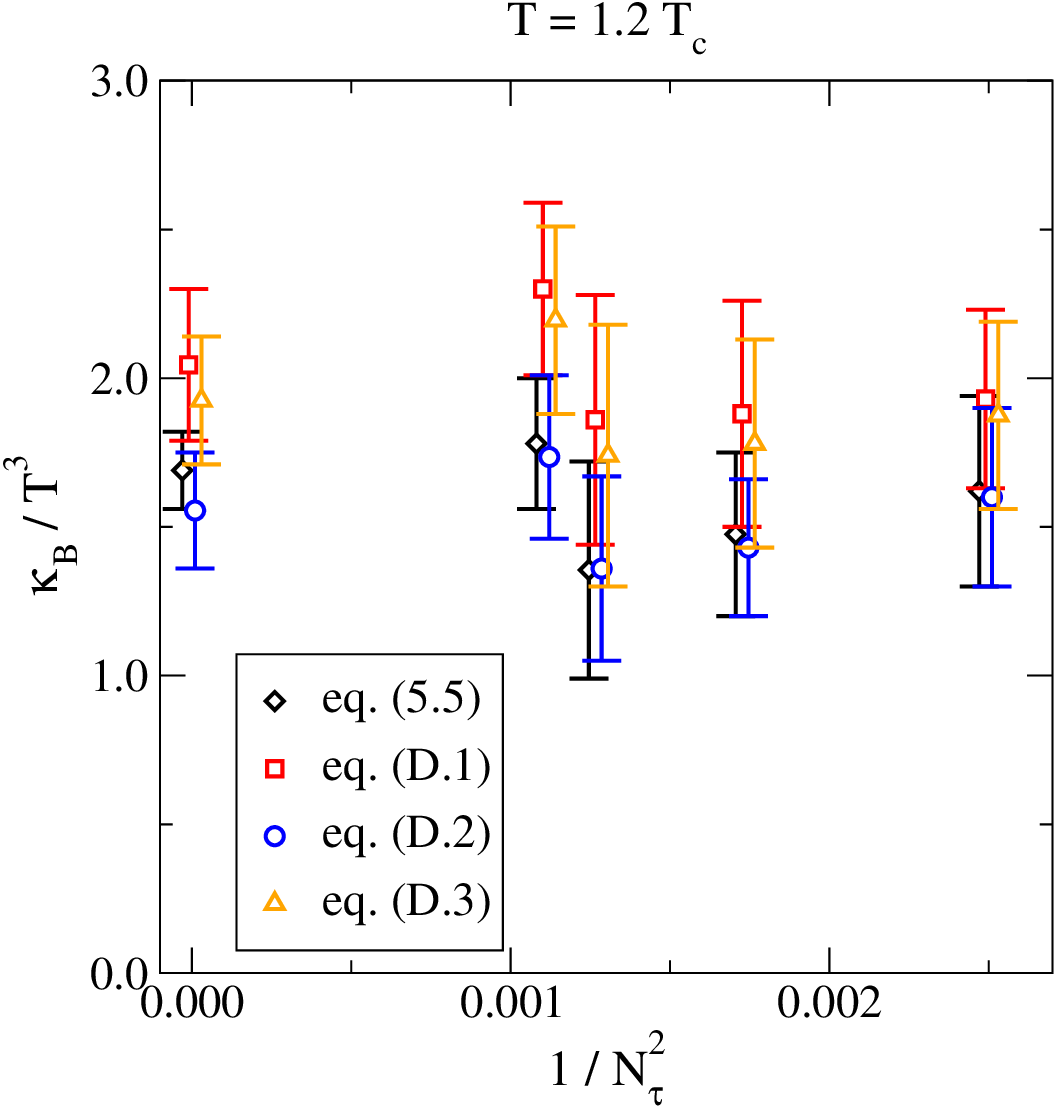}
 ~~\epsfxsize=5.0cm\epsfbox{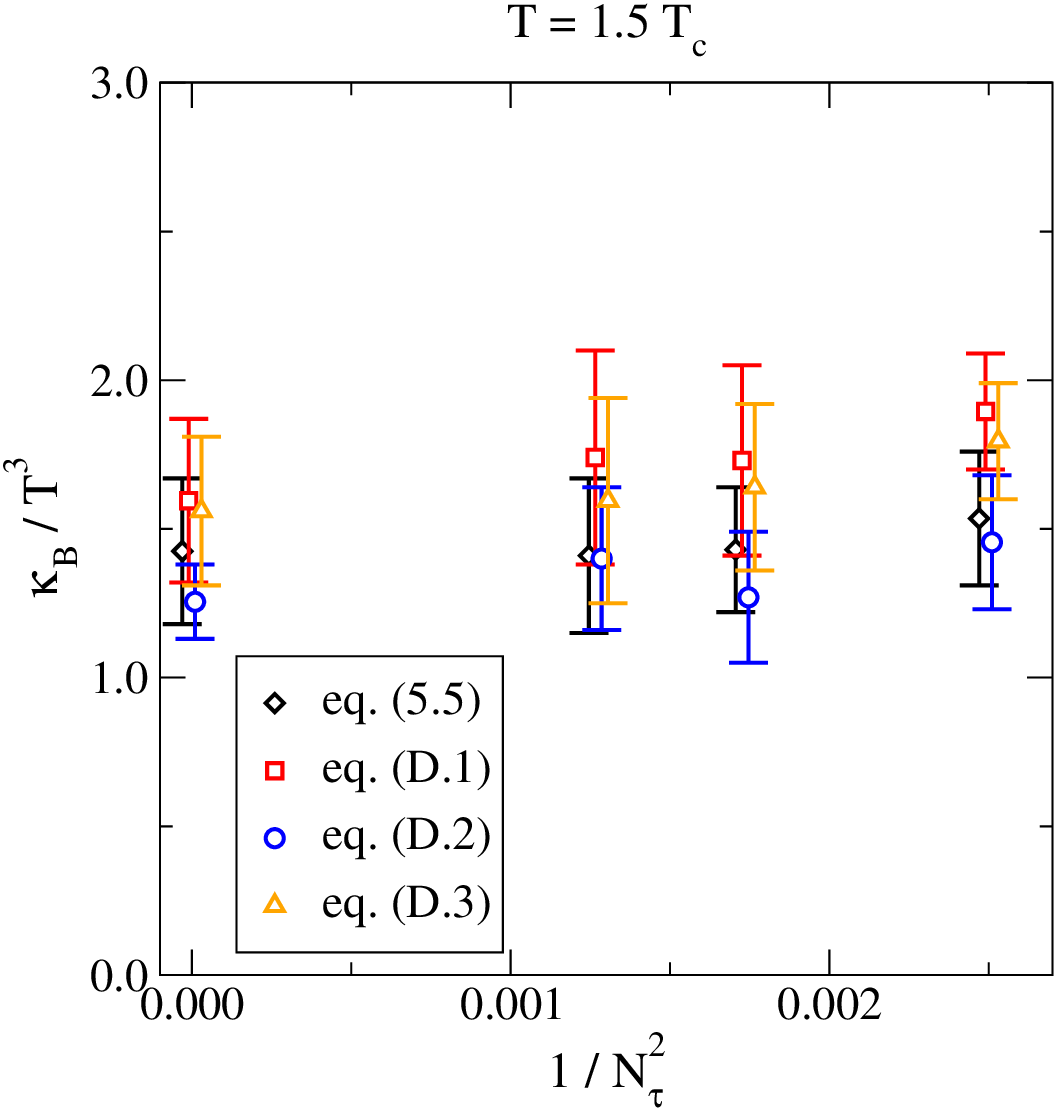}
 ~~\epsfxsize=5.0cm\epsfbox{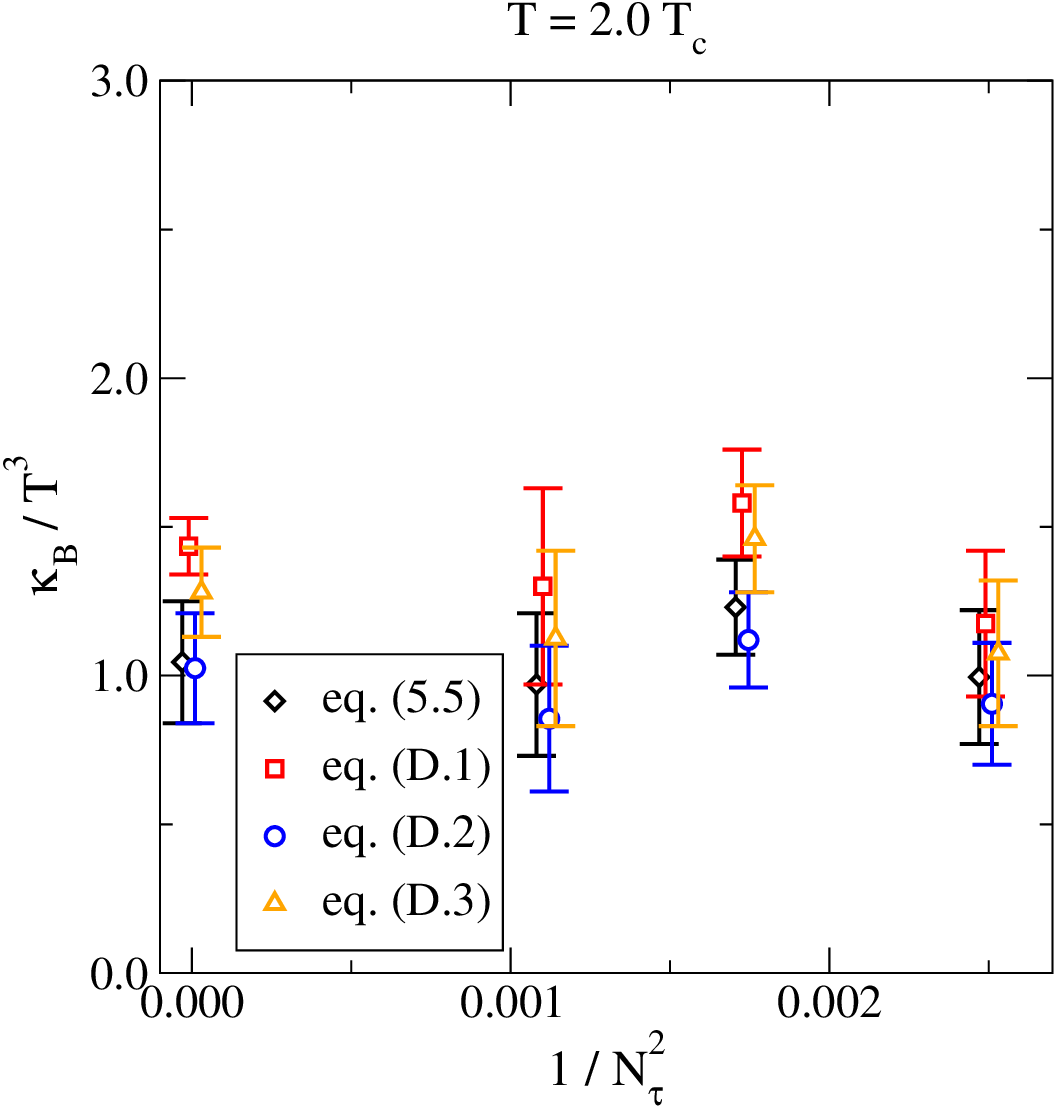}
}

\caption[a]{\small
     The parameter $\kappa^{ }_{\B} / T^3$ as obtained from
     the different fits, according to  
     tables~\ref{table:1.2Tc}--\ref{table:2.0Tc}. 
     For better visibility, 
     the data have been slightly displaced in the horizontal direction. 
}

\la{fig:fits}
\end{figure}

%
\begin{table}[t]

\small{
\begin{center}
\begin{tabular}{ccccc} 
 fit form & 
 $N^{ }_\tau = 20$ &  
 $N^{ }_\tau = 24$ &  
 $N^{ }_\tau = 30$ &  
 $N^{ }_\tau\to \infty $
 \\[2mm]
 \hline
 \\[-4mm] 
 \eq\nr{model_B} &
      0.77 -- 1.22 & 1.07 -- 1.39
    & 0.73 -- 1.21 & 0.84 -- 1.25
  \\[0.5mm]  
 \eq\nr{formc} &
      0.93 -- 1.42 & 1.40 -- 1.76
    & 0.97 -- 1.63 & 1.34 -- 1.53
 \\[0.5mm]  
 \eq\nr{fourierb} & 
      0.70 -- 1.11 & 0.96 -- 1.28
    & 0.61 -- 1.10 & 0.84 -- 1.21
 \\[0.5mm]  
 \eq\nr{fourierc} &
      0.83 -- 1.32 & 1.28 -- 1.64
    & 0.83 -- 1.42 & 1.13 -- 1.43
 \\  
 \hline 
\end{tabular} 
\end{center}
}


\caption[a]{\small
 Fit results for $\kappa^{ }_{\B}/T^3$ 
 at $T = 2.0 \Tc^{ }$, 
 as described in appendix~D.  
 }
\label{table:2.0Tc}
\end{table}
%

We give here more details on the fitting procedure employed 
in \se\ref{se:implications}, which in the end 
produced the results shown in \fig\ref{fig:results}, 
after combining all errors in the most conservative way
(from absolute minimum among the different fit forms, 
to the absolute maximum).

Four different forms have been employed for the spectral function: 
\eq\nr{model_B}, as well as 
\ba
 \rho^\rmii{\nr{formc}}_{\B}(\omega) & \equiv &
 \max\{ {\phi}^{ }_\rmii{IR}(\omega), 
    b^{ }_{\B \rmii{1}}\, \phi^{ }_\rmii{UV}(\omega) 
 \}
 \;, \la{formc} \\
 \rho^\rmii{\nr{fourierb}}_{\B}(\omega) & \equiv &
 \bigl( 1 + b^{ }_{\B \rmii{2}} \sin\pi y \bigr)
  \sqrt{
    {\phi}^{2}_\rmii{IR}(\omega) \; + \; 
     \phi^{2}_\rmii{UV}(\omega) 
 }
 \;, \la{fourierb} \\
 \rho^\rmii{\nr{fourierc}}_{\B}(\omega) & \equiv &
 \bigl( 1 + b^{ }_{\B \rmii{3}} \sin\pi y \bigr)
 \max\{ {\phi}^{ }_\rmii{IR}(\omega), 
         \phi^{ }_\rmii{UV}(\omega) 
 \}
 \;. \la{fourierc} 
\ea
In the last two cases, following ref.~\cite{lat4}, we have defined 
\be
 x \; \equiv \; \ln\Bigl( 1 + \frac{\omega}{\pi T} \Bigr) 
 \;, \quad
 y \; \equiv \; \frac{x}{1+x} 
 \;, 
\ee
thereby mapping the $\omega$-range $(0,\infty)$ 
onto the interval $(0,1)$ such 
that at both ends the $\omega$-dependence has a qualitatively
reasonable form. The auxiliary fit parameters 
$a^{ }_\B$, $b^{ }_{\B \rmii{1}}$ are of order unity, 
whereas $b^{ }_{\B \rmii{2}}$, 
$b^{ }_{\B \rmii{3}}$ are in the range $0.02-0.10$. 
Our main
interest is in the fit parameter $\kappa^{ }_{\B}/T^3$, 
which enters through  $ {\phi}^{ }_\rmii{IR} $
(cf.\ \eq\nr{phiIR}).

Since the continuum extrapolation, 
as described in \se\ref{se:data},  
involves various interpolations, with their own sets of uncertainties,  
we fit separately to the correlators at finite $N^{ }_\tau$,
in addition to the continuum-extrapolated correlators. 

At $1.2 \Tc^{ }$, we have four lattice spacings.
The results are shown in table~\ref{table:1.2Tc} and
in the left panel of \fig\ref{fig:fits}.
In each case the fit was
carried out for $\bar\tau \in (\tlo,0.5/T)$, 
where $\tlo$ was varied in the range
$\sim (0.25 - 0.35)/T$ for finite $N^{ }_\tau$ 
and $\tlo \sim 0.17/T$ for
the continuum-extrapolated results. 
The error was obtained from a bootstrap analysis. 
The median of the bootstrap distribution of $\chi^2/$d.o.f.\ 
was comfortably $ < 1.0$,
and always below $\sim 1.5$.
The error bands include, besides this bootstrap error, 
the spread due to changing $\tlo$. 

At $1.5 \Tc^{ }$ and $2.0 \Tc^{ }$, we have three lattice spacings. 
The results are shown in tables~\ref{table:1.5Tc} 
and~\ref{table:2.0Tc} and in the two right-most panels
of \fig\ref{fig:fits}, respectively.
In each case the fit was carried out for 
$\bar\tau \in (\tlo,0.5/T)$, where $\tlo$ was varied in the range
$\sim (0.20 - 0.35)/T$, where $\chi^2/$d.o.f.\ was acceptable. 
The interpretation of the error bands is similar to that at $1.2\Tc^{ }$.

\small{
%

}

\end{document}